\documentclass[12pt,notitlepage]{article}

\usepackage{amsmath,amssymb,amsfonts,amsthm,color,fancybox,wrapfig,url,paralist,float,verbatim,varioref,overpic,wrapfig}
\usepackage{graphicx}
\usepackage{geometry,setspace,times,ifthen}
\usepackage{fancyhdr}
\usepackage[ruled,vlined]{algorithm2e}
\usepackage{caption}
\usepackage{subcaption}

\usepackage{tikz}
\usepackage[european resistor, european voltage, european current]{circuitikz}
\usetikzlibrary{arrows,shapes,positioning,snakes,calc}
\usetikzlibrary{decorations.markings,decorations.pathmorphing,decorations.pathreplacing}
\usetikzlibrary{calc,patterns,shapes.geometric,topaths,fit}

\renewcommand{\thepage}{\arabic{page}}

\fancypagestyle{secondstyle}
{
\renewcommand{\thepage}{\arabic{page}}
\lhead{\sffamily Vikram Krishnamurthy}
\chead{\today}
\rhead{\sffamily \thepage}

\lfoot{}
\cfoot{}
\rfoot{}
}

\usepackage{mathptmx}

\newcommand{\titlename}{Adversarial Radar Inference. From Inverse Tracking  to Inverse Reinforcement Learning of Cognitive Radar}

\makeatletter
\def\nl#1#2{\begingroup
    #2%
    \def\@currentlabel{#2}%
    \phantomsection\label{#1}\endgroup
}
\makeatother

{\ensuremath{
\begin{empheq}[box=\fbox]{align}
{#1}
\end{empheq}
}}

\newtheorem{theorem}            {Theorem}[section]

\newtheorem{definition}         [theorem]{Definition}

\newtheorem{lemma}              [theorem]{Lemma}

\newcommand{\normal}{\mathbf{N}}  

{
\begin{mdframed}
\par\noindent\textbf{#1:}\begin{rmfamily}\noindent}%
{\end{rmfamily}
\end{mdframed}
}

\newcommand{\hobsm}{\obsm}
\newcommand{\post}{\alpha}

\newcommand{\aprob}{G}

\newcommand{\unpost}{q}
\newcommand{\unpostm}{\unpost^\model}
\newcommand{\horizon}{N}

\newcommand{\kg}{\psi}

\newcommand{\obsmt}{C^o}

\newcommand{\anoise}{\epsilon}

\newcommand{\pdf}{p}

\newcommand{\prob}{\mathbb{P}}
\newcommand{\E}                 {\Bbb{E}}

\newcommand{\obs}{y}

\newcommand{\snoise}{w}
\newcommand{\onoise}{v}
\newcommand{\statem}{A}

\newcommand{\obsm}{C}

\newcommand{\snoisecov}{Q}
\newcommand{\onoisecov}{R}

\newcommand{\state}{x}
\newcommand{\statespace}{\mathcal{X}}
\newcommand{\obspace}{\mathcal{Y}}
\newcommand{\statedim}{X}
\newcommand{\obsdim}{{Y}}

\newcommand{\fun}{\phi}

\newcommand{\oprob}{B}
\newcommand{\tp}{P}

\newcommand{\finaltime}{N}

\newcommand{\model}{\theta}

\newcommand{\Model}{\Theta}
\newcommand{\lik}{L_\finaltime} 

\newcommand{\probe}{\alpha}
\newcommand{\Probe}{\boldsymbol{\probe}}
\newcommand{\response}{\beta}
\newcommand{\dtime}{n}

\newcommand{\mle}{\model^*}

\newcommand{\belief}{\pi}

\newcommand{\Belief}{\Pi}

\newcommand{\modeldim}{M}

\newcommand{\eobs}{z}

\newcommand{\enemystate}{\hat{\hat{\state}}}
\newcommand{\enemystatem}{\bar{\statem}}
\newcommand{\enemyobsm}{\bar{\obsm}}
\newcommand{\enemykalmancov}{\bar{\kalmancov}}

\newcommand{\enemySig}{\bar{\Sig}}
\newcommand{\enemyonoisecov}{\bar{\onoisecov}}
\newcommand{\enemysnoisecov}{\bar{\snoisecov}}
\newcommand{\enemyinputm}{\bar{F}}
\newcommand{\enemykg}{\bar{\kg}}

\newcommand{\eaction}{\bar{a}}

\makeatletter
\newcommand{\pushright}[1]{\ifmeasuring@#1\else\omit\hfill$\displaystyle#1$\fi\ignorespaces}
\makeatother

  {\popQED\end{theorem}}

\newcommand{\anoisecov}{\sigma^2_\anoise}

   \newcommand{\laction}{u}

\newcommand{\kalmancov}{\Sigma}

\newcommand{\Sig}{S}   

\newcommand{\imp }{\pi}

\newcommand{\ole}{\stackrel{\text{defn}}{=}}

\newcommand{\filter}{T}

\newcommand{\argmin}{\operatornamewithlimits{argmin}}
\newcommand{\argmax}{\operatornamewithlimits{argmax}}

\newcommand{\reals}{{\rm I\hspace{-.07cm}R}}

\newcommand{\beq}{\begin{equation}}
\newcommand{\eeq}{\end{equation}}

\newcommand{\p}{\prime}

\newcommand{\action}{a}

\newcommand{\actiondim}{U}

\def\param{{\alpha}}

\newcommand{\innovations}{\iota}

\newcommand{\threshold}{\gamma}

\newcommand{\nresponse}{\obsresponse}

\newcommand{\bound}{\bar{\kalmancov}}

\newcommand{\ARE}{\operatorname{\mathcal{A}}}

\newcommand{\obsdataset}{\mathcal{D}_\text{obs}}

\newcommand{\hstate}{\hat{\state}}

\newcommand{\barray}{\begin{array}{ll}}
\newcommand{\earray}{\end{array}}

\newcommand{\Response}{\boldsymbol{\response}}
\newcommand{\Anoise}{\boldsymbol{\anoise}}

\newcommand{\lagrange}{\lambda}
\newcommand{\probedim}{m}

\newcommand{\sindx}{s}
\newcommand{\tindx}{t}

\newcommand{\utility}{U}

\newcommand{\dataset}{\mathcal{D}}

\newcommand{\obsresponse}{z}

\usepackage{quoting}
\quotingsetup{vskip=1pt}

\usepackage[labeled,resetlabels]{multibib}

\newcites{A}{{Bibliography:  Related Work by the Principal Investigator}}

\tikzset{
    block/.style={rectangle, draw, line width=0.5mm, black, text width=4.5em, text centered,
                 minimum height=1em},
               line/.style={draw, -latex}}

      \tikzset{
    block3/.style={rectangle, draw, line width=0.5mm, black, text width=7.5em, text centered,
                 minimum height=1em},
               line/.style={draw, -latex}}

\tikzset{
    block2/.style={rectangle, draw, line width=0.5mm, text centered,
                 minimum height=2em},
               line/.style={draw, -latex}}

\tikzset{
    blocka/.style={rectangle, draw, line width=0.5mm, black, text width=4.5em, text centered,
                 minimum height=1em},
               line/.style={draw, -latex}}

\begin{document}

\pagestyle{secondstyle}

 \setcounter{page}{1}
\begin{center} {\Large {\bf  \titlename}} \\ \vspace{0.5cm}
{ Vikram Krishnamurthy, School of Electrical \& Computer Engineering, Cornell University, vikramk@cornell.edu}
\end{center}


\section{Introduction}

Cognitive
sensing  refers to a reconfigurable sensor that dynamically adapts its sensing
mechanism  by using  stochastic  control to  optimize its sensing resources. For example, cognitive radars  are sophisticated dynamical systems; they  use
stochastic control  to  sense the environment, learn fro<m it relevant information about the target and  background, then adapt the radar sensor to  satisfy the needs of their mission.  The last two decades have witnessed intense research in cognitive/adaptive radars ~\cite{CKH09,KD07,KD09,Hay12}.

This chapter discusses   addresses the  next logical step, namely {\em inverse cognitive sensing}.
 By observing the emissions of  a  sensor (e.g.\  radar or in general a controlled stochastic dynamical system) in real time, how can we detect if the sensor  is cognitive (rational utility maximizer) and how can we predict its future actions?
The scientific challenges  involve extending Bayesian filtering, inverse reinforcement learning  and stochastic optimization  of  dynamical systems  to a data-driven adversarial setting. 
Our methodology transcends  
classical 
statistical signal processing (sensing and estimation/detection theory) to
address the deeper issue of {\em how to infer strategy from sensing}.
The generative models,  adversarial inference  algorithms and associated mathematical analysis will lead to advances  in understanding how sophisticated adaptive sensors such as cognitive radars operate. 
As will be discussed below, the framework involves the interaction of two DDDAS (dynamic data driven applications systems).

   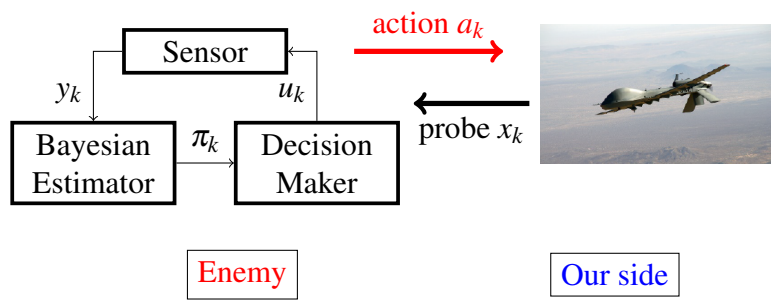
\begin{figure}[h] \centering
      {\resizebox{9cm}{!}{
\begin{tikzpicture}[node distance = 1cm, auto]
    \node [block] (BLOCK1) {Sensor};
    \node [block, below of=BLOCK1,right of=BLOCK1,node distance=1.5cm] (BLOCK2) {Decision \\ Maker};
    \node [block, below of=BLOCK1,left of=BLOCK1,node distance=1.5cm] (BLOCK3) {Bayesian Estimator};

    \draw[<-] (BLOCK1) -| node[left,pos=0.8]{$\laction_k$}  (BLOCK2)  ;
    \draw[->] (BLOCK1.west) -|   node[left,pos=0.8]{$\obs_k$} (BLOCK3);

    \draw[->](BLOCK3) --  node[above]{$\belief_k$} (BLOCK2);

    \node[draw=none,fill=none] at (4.5,-1.5) (drone) {\includegraphics[bb=0 0 0 0,scale=0.4]{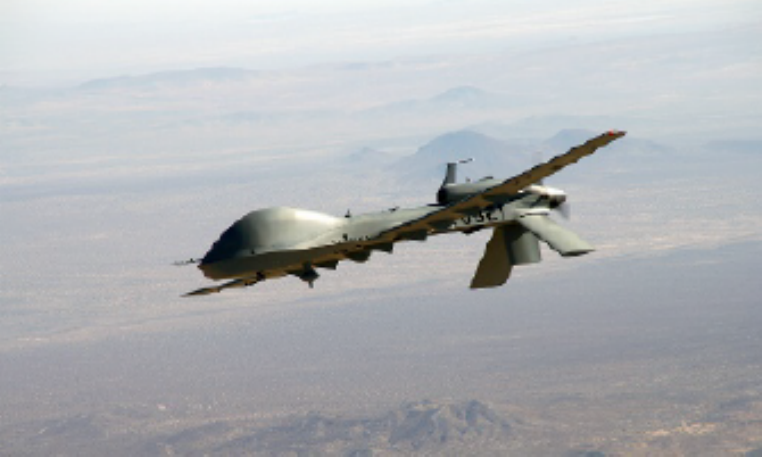}};

    \draw[->,color=red,line width=2pt] (2,0)   -- node[above]{action $\action_k$}(4,0);
    \draw[->,line width=2pt] ([yshift=0.8cm]drone.west)   --   node[below]{probe $\state_k$} (2.8,-0.7);
    \node[draw] at (5.5,-3) {{\color{blue}Our side}};
    \node[draw] at (0.5,-3) {{\color{red}Enemy}};
    

  \end{tikzpicture}} }

\caption{\small \textbf{Schematic of Adversarial Inference Problem.} ``Our'' side is a drone/UAV or electromagnetic signal that probes the  enemy's cognitive sensor (e.g.\ radar). Given our state sequence $\{x_k\}$ and observed enemy's actions $\{a_k$\}, our aims are to: (i) Estimate the enemy's belief (posterior) $\belief_k$ and its sensor accuracy (ii) Optimally probe the enemy with $\{\state_k\}$ to minimize the variance of our estimate of its accuracy \\  (iii)  Determine if the enemy is cognitive  (rational utility maximizer), and  estimate its utility function  \\ (iv)  Track time varying equilibria in a game theoretic setting
} \label{fig:schematic}
\end{figure}

\subsection{Objectives} \label{sec:objectives}
   The central theme involves an adversarial signal processing problem  comprised of   ``us'' and an ``adversary''.
``Us'' refers to a drone/UAV  or electromagnetic signal that probes an ``adversary'' cognitive  sensing  system.
 Figure~\ref{fig:schematic} shows the schematic setup. A cognitive sensor  observes 
  our kinematic state $\state_k$ in noise as the observation $\obs_k$. It then  uses a Bayesian tracker to update its posterior distribution $\belief_k$ of our state $\state_k$ and chooses an action $\laction_k$ based on this posterior. We observe the sensor's action in noise as $\action_k$. 
Given knowledge of ``our'' state sequence  $\{\state_k\}$  and the observed    actions $\{\action_k\}$ taken by the enemy's sensor, 
this  project focuses on the following   inter-related aspects; each topic below is  backed by extensive preliminary research and results:

 {\bf  1. Inverse filtering and Estimating the Enemy's Sensor Gain}. How to estimate the enemy's   estimate of us?
  Suppose the enemy observes our  state in noise; updates its posterior  distribution $\belief_k$ of our state $\state_k$ and then chooses an action $\laction_k$ based on this posterior. Given knowledge of ``our'' state and sequence of  enemy's actions observed in noise $\{\action_k\}$, how can the enemy's posterior distribution (random measure) be estimated? We will  develop computationally efficient filtering algorithms with performance bounds.

A related question is:
How to remotely estimate  the enemy's  sensor observation likelihood when it  is estimating us?  This is important because it tells us how accurate the enemy's sensor is; in the context of Figure \ref{fig:schematic} it tells us, how accurately the enemy tracks our drone. The data we have access to is our
 state (probe signal) $\{\state_k\}$ and  measurements of the enemy's radar actions $\{\action_k\}$.
Estimating the enemy's sensor accuracy is  non-trivial with several challenges.
   First, even though we know our state and state dynamics model (transition law),
 the enemy does not.
 The enemy  needs to estimate our state and state transition law  based on our trajectory; and we  need to estimate the enemy's estimate of our
 state transition law. Second, computing the maximum likelihood estimate of the enemy's sensor gain involves inverse filtering;  as shown in our preliminary work, the convergence rate of numerical algorithms to compute the estimate is substantially slower than classical maximum likelihood estimation. We will also study identifiability issues and consistency properties of the resulting maximum likelihood estimator.

{\bf 2. Optimal Probing of Adversary}.  Referring  to Figure \ref{fig:schematic}, how can we optimally probe the enemy's sensor   to minimize the covariance of our estimate of the enemy's  observation likelihood? ``Our'' state can be viewed as a probe signal which causes the adversary's radar  to act; so choosing the optimal state sequence is an input/experimental design problem. 
We will consider two approaches. The first  approach is analytical -- using stochastic dominance, we will establish a partial ordering amongst probe transition  matrices which results in a  corresponding ordering of our SNR of the adversary's actions. 
The second approach is  numerical -- we will construct  stochastic gradient algorithms to optimize the probe transition matrix.

  {\bf 3. Revealed Preferences and Inverse Reinforcement Learning.} 
 Our working assumption is that a cognitive sensor  satisfies economic rationality, i.e., a cognitive sensor is a utility maximizer that optimizes its actions $\action_k$ subject to physical level (Bayesian filter) constraints.
How can we detect this utility maximization behavior?
 Nonparametric detection of utility maximization behavior is the central theme of {\em revealed preferences} in microeconomics. A remarkable result is \textit{Afriat's theorem}: it  provides a necessary and sufficient condition for a finite dataset  to
 have originated from a utility maximizer. 
 We will  develop constrained utility estimation methods in a Bayesian framework that  account for signal processing constraints introduced  by 
the Bayesian tracker. These impose  nonlinear constraints involving the spectrum (eigenvalues) of the achieved covariance of the tracker; such spectral revealed preferences require  generalizing  the classical Afriat theorem.

When the enemy's actions are observed in noise, how can a statistical detector be constructed to test for utility maximization behavior?
We will also address the related issue of: 
How can our state (probe signal) be chosen by us  to minimize the Type 2  error of detecting if the adversary is deploying an economic rational strategy, subject to a constraint on the Type 1 detection error?

\subsection{Perspective and Context. Relevance to DDDAS}
The adversarial dynamics considered in this chapter fit naturally within the  \textit{Dynamic Data Driven Applications Systems  (DDDAS)} paradigm. 
 The adversary's radar  senses, adapts and learns from us. In turn we adapt sense and learn from the adversary. So in simple terms we are modeling and analyzing the interaction of two DDDAS. In this context this chapter has two major themes: inverse filtering which is a Bayesian framework for interacting DDDAS, and inverse cognitive sensing which is a non-parametric approach for utility estimation for interacting DDDAS.

Here are some  unusual features that motivate the
ideas in this chapter:
  \begin{compactenum}[(i)]
    \item Slow Learning: Inverse filtering (estimating the enemy's estimate of us) has slow convergence - given $n$ data points the covariance decreases as $O(n^{-1/3})$, whereas the classical Kalman filter covariance decreases as $O(n^{-1})$. The change in convergence speed exhibits a remarkable phase transition.
  \item  Data driven approach:
 The revealed preference approach  is  fundamentally different to the {\em model-centric} theme that is widely  used in signal processing   where
one postulates an objective function (typically convex) and then proposes  optimization algorithms.  In contrast,  the revealed
preference approach is {\em data centric} - given a dataset,  we wish to determine if is consistent with utility maximization, i.e., if the sensor is satisfies economics-based rationality.
\item Classical inverse reinforcement learning aims to estimate the utility function
of a decision maker by observing its decisions and  input state
\cite{NR00};
 the
existence of a utility function (rationality) is assumed implicitly. The revealed preference approach considered in this chapter
addresses a deeper and more fundamental question: does a utility function exist that rationalizes the
given data (with  signal processing constraints) and if yes, estimate it. As discussed  below, inverse reinforcement learning is  a trivial case of revealed preferences.

\item  Finally, this chapter   is an early step  in understanding
how to design stealthy cognitive sensors whose cognitive functionality is difficult to detect by an observer. (\textit{In simple terms, how to design a  smart sensor that acts dumb?}). This  generalizes the physics based low-probability of intercept (LPI) requirement of radar (which requires low power emission) to the    systems-level issue: How should the sensor choose its actions in order to avoid detection of its cognition?

 This work is also  motivated by the design of counter-autonomous systems: given measurements of  the actions of an  autonomous adversary, how can our counter-autonomous system  estimate the underlying belief of the adversary,  predict future actions and therefore guard against these actions. 
Specifically, \cite{Kup17} places counter unmanned autonomous systems at a level of abstraction above the physical sensors/actuators/weapons and datalink layers; and below the human controller layer.
\end{compactenum}

\section{Inverse Filtering and Estimating Enemy's Sensor}  \label{sec:cspomdp}

This section discusses   inverse filtering in an adversarial system, see Figure \ref{fig:graph} for the schematic setup. Our main ideas involve   estimating the enemy's estimate of us and estimating the enemy's sensor observation likelihood.

\subsection{Background and Preliminary Work}

    \begin{figure}[h]
      \begin{subfigure}{0.47\textwidth}
      {\resizebox{8cm}{!}{
\begin{tikzpicture}[node distance = 1cm, auto]
    \node [block] (BLOCK1) {Sensor};
    \node [block, below of=BLOCK1,right of=BLOCK1,node distance=1.5cm] (BLOCK2) {Decision \\ Maker};
    \node [block, below of=BLOCK1,left of=BLOCK1,node distance=1.5cm] (BLOCK3) {Tracker $\filter(\belief_{k-1},\obs_k)$};

    \draw[<-] (BLOCK1) -| node[left,pos=0.8]{$\laction_k$}  (BLOCK2)  ;
    \draw[->] (BLOCK1.west) -|   node[left,pos=0.8]{$\obs_k$} (BLOCK3);

    \draw[->](BLOCK3) --  node[above]{$\belief_k$} (BLOCK2);

    \node[draw=none,fill=none] at (4.5,-1.5) (drone) {\includegraphics[bb=0 0 0 0,scale=0.4]{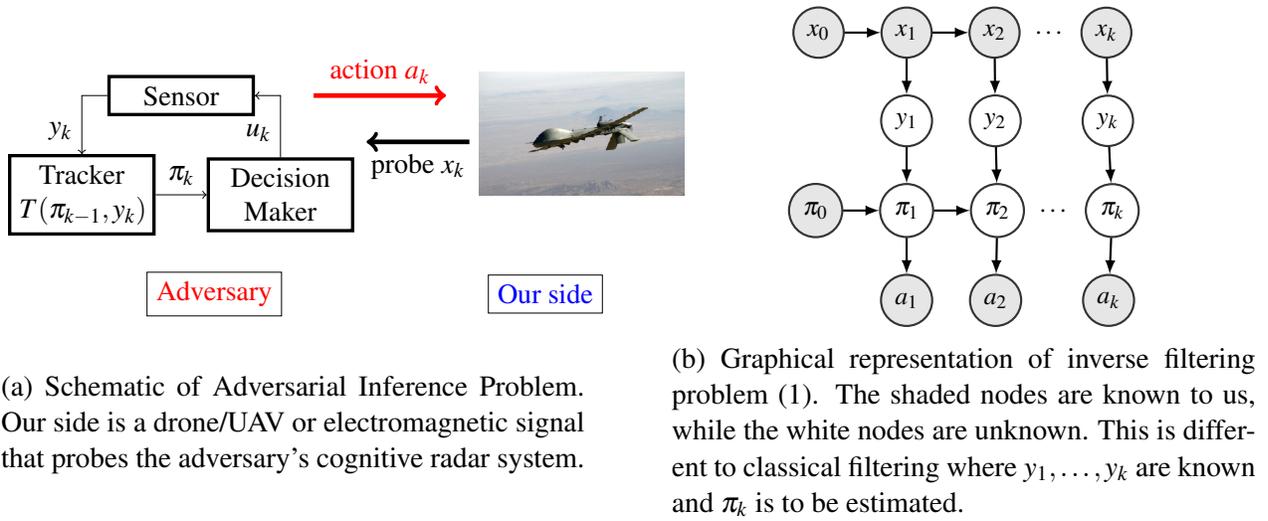}};

    \draw[->,color=red,line width=2pt] (2,0)   -- node[above]{action $\action_k$}(4,0);
    \draw[->,line width=2pt] ([yshift=0.8cm]drone.west)   --   node[below]{probe $\state_k$} (2.8,-0.7);
    \node[draw] at (5.5,-3.0) {{\color{blue}Our side}};
    \node[draw] at (0.5,-3.0) {{\color{red}Adversary}};
    

  \end{tikzpicture}} }

\caption{Schematic of Adversarial Inference Problem. Our side is a drone/UAV or electromagnetic signal that probes the  adversary's cognitive radar system.}
\end{subfigure}
\hspace{1cm}
\begin{subfigure}{0.47\textwidth}  \centering
{
\begin{tikzpicture}[thick,scale=0.8, every node/.style={transform shape}]
\tikzstyle{main}=[circle, minimum size = 5mm, thick, draw =black!80, node distance = 6mm]
\tikzstyle{connect}=[-latex, thick]
  \node[main, fill = black!10] (x0) {$x_0$};
  \node[main, fill = black!10] (x1) [right=of x0] {$x_1$};
  \node[main, fill = black!10] (x2) [right=of x1] {$x_2$};
  \node[main, fill = black!10] (xk) [right=10mm of x2] {$x_k$};
  \path (x0) edge [connect] (x1)
        (x1) edge [connect] (x2)
        (x2) -- node[auto=false]{\ldots} (xk);
  
  \node[main] (y1) [below=of x1] {$y_1$ };
  \node[main] (y2) [right=of y1] {$y_2$};
  \node[main] (yk) [right=10mm of y2] {$y_k$};
  \path (x1) edge [connect] (y1)
        (x2) edge [connect] (y2)
        (xk) edge [connect] (yk);
  
  \node[main] (pi1) [below=of y1] {$\pi_1$ };
  \node[main, fill = black!10] (pi0) [left=of pi1]{$\pi_0$};
  \node[main] (pi2) [right=of pi1] {$\pi_2$};
  \node[main] (pik) [right=10mm of pi2] {$\pi_k$};
  \path (pi0) edge [connect] (pi1)
        (pi1) edge [connect] (pi2)
        (pi2) -- node[auto=false]{\ldots} (pik)
    
        (y1) edge [connect] (pi1)
        (y2) edge [connect] (pi2)
        (yk) edge [connect] (pik);

  \node[main, fill = black!10] (a1) [below=of pi1]{$a_1$};
  \node[main, fill = black!10] (a2) [right=of a1] {$a_2$};
  \node[main, fill = black!10] (ak) [right=10mm of a2] {$a_k$};
  \path (pi1) edge [connect] (a1)
        (pi2) edge [connect] (a2)
        (pik) edge [connect] (ak);
      \end{tikzpicture}}
    \caption{Graphical representation of inverse filtering problem (\ref{eq:model}). The shaded nodes are known to us, while the white nodes are unknown. This is different to classical filtering where $\obs_1,\ldots,\obs_k$ are known and $\belief_k$ is to be estimated.}
    \end{subfigure}
    \caption{Schematic and Graphical representation of adversarial inverse filtering  problem considered in this chapter.}
\label{fig:graph}
\end{figure}

The problem formulation involves two players;  ``us'' and ``adversary''.  With $k=1,2,\ldots$ denoting discrete time, the model has
the following dynamics:
\begin{equation}
\begin{split}
    \state_k &\sim  \tp_{\state_{k-1},\state} = \pdf(\state | \state_{k-1}), \quad \state_0 \sim \belief_0 \\
    \obs_k  &\sim \oprob_{\state_k,\obs} = \pdf(y | x_k)\\
    \belief_k &= \filter(\belief_{k-1}, \obs_k)\\
    \action_k &\sim \aprob_{\belief_k,\action} = \pdf(\action | \belief_k)
  \end{split} \label{eq:model}
\end{equation}
Let us explain the notation in (\ref{eq:model}):
\begin{compactitem}
    \item $\state_k\in \statespace$ is our Markovian state with  transition kernel $\tp_{\state_{k-1},\state}$,  prior $\belief_0$ and state space  $\statespace$.
    \item $\obs_k$ is the adversary's noisy observation of our state $\state_k$; with observation likelihoods $\oprob_{xy}$.
    \item $\belief_k = \pdf(x_k| \obs_{1:k})$ is the adversary's belief (posterior)  of our state $\state_k$ where $\obs_{1:k}$ denotes the sequence  $\obs_1,\ldots,\obs_k$. The operator $T$ in (\ref{eq:model}) is the classical Bayesian optimal  filter
       \beq  \filter(\belief,\obs) (\state) = \frac{
    \oprob_{\state ,\obs} \int_\statespace 
    \tp_{\zeta, \state}\, \belief(\zeta) \,d\zeta}
  {\int_\statespace  \oprob_{\state ,\obs} \int_\statespace 
    \tp_{\zeta, \state}\, \belief(\zeta)\, d\zeta d\state}
  \label{eq:belief}
\eeq
Let $\Belief$ denote the  space of all such beliefs. When
the state space
$\statespace$ is finite, then $\Belief$ is  the unit $
\statedim-1$ dimensional simplex of $\statedim$-dimensional probability mass functions.
    \item $\action_k $ denotes our measurement of the adversary's action based on its current belief $\belief_k$. More explicitly, the adversary chooses an action $\laction_k$ as a deterministic function of $\belief_k$ and we observe $\laction_k$ in noise as $\action_k$. We encode this as $\aprob_{\belief_k,\action_k}$; this is the conditional probability of observing  action $\action_k$ given the adversary's belief $\belief_k$.
    \end{compactitem}

Figure \ref{fig:graph} displays a schematic and graphical representation of the model (\ref{eq:model}). The schematic model shows ``us'' and the adversary's variables. In the graphical model, the shaded nodes are known to us, while the white nodes are computed by the adversary and unknown to us. This is in contrast to classical  filtering where $\obs_k$ is known to us and $\state_k$ is to be estimated, i.e, $\belief_k$ is to be computed.
\\
{\bf Aim}: Referring to model (\ref{eq:model}) and  Figure \ref{fig:graph}, our aims are:
\begin{compactenum} \item How to estimate the adversary's belief given measurements of its actions (which are based on  its filtered estimate of our state)? In other words, assuming  probability distributions  $\tp,\oprob,\aprob$ are  known, we aim to estimate  the adversary's belief $\belief_k$ at each time $k$, by computing   posterior $\pdf(\belief_k\mid \belief_0,\state_{0:k},\action_{1:k})$. 
\item How to estimate the enemy's observation kernel $\oprob$, i.e its sensor gain? This  tells us how accurate the enemy's sensor is.
\end{compactenum}
From a practical point of view, estimating the enemy's belief and sensor parameters allows us to  calibrate its accuracy and predict (in a Bayesian sense)  future actions of the enemy.
\\ 
{\bf Related  Works}.
In our recent works  \cite{MRKW17,MRKW18,MIC19}, the mapping from
belief $\belief$ to enemy's action $\action$ was assumed  deterministic. Specifically, \cite{MRKW17} gives a deterministic regression based approach to  estimate the adversary's model parameters in a Hidden Markov model. In comparison, our proposed research here  assumes a probabilistic map between $\belief $ and $\action$ and we  develop Bayesian filtering algorithms for estimating  the posterior along with MLE algorithms for $\model$.
Estimating/reconstructing the posterior given decisions based on the posterior  is  studied in microeconomics under the area of
   social learning \cite{Cha04,MTJ18} and game-theoretic  generalizations 
\cite{AHP07}. 
There  are strong parallels between inverse filtering and Bayesian social learning  \cite{Cha04}, \cite{Kri16,Kri12,Kri11}; the key difference is that social learning aims to estimate the underlying state given noisy posteriors, whereas our aim is to estimate the posterior given noisy measurements of the posterior and the underlying state. Recently,  \cite{HAB18} used cascaded Kalman filters for LQG  control over  communication channels.

\subsection{Inverse Filtering Algorithms}
\begin{quote}
{\bf {\em   How to estimate the enemy's posterior distribution $\belief_k$ of us?}}
\end{quote}
Here we discuss inverse filtering for special cases of the model (\ref{eq:model}).
Define the posterior distribution $\post_{k}(\belief_k) = \pdf(\belief_k |\action_{1:k},\state_{0:k})$ of the adversary's posterior
distribution given our state sequence $\state_{0:k}$ and actions $\action_{1:k}$:
Note that the posterior $\post_k(\cdot)$ is a {\em random measure} since it is a posterior distribution of the adversary's posterior  distribution (belief) $\belief_k$.
By using a discrete time version of Girsanov's theorem and appropriate change of measure\footnote{This chapter deals with discrete time. Although we will not pursue it here, our recent paper \cite{KLS18} uses a similar  continuous time formulation. This  yields interesting results involving Malliavin derivatives  and stochastic calculus.} \cite{EAM95} (or  a careful application of Bayes rule) we can derive the following functional recursion for $\post_k$:
\beq
  \post_{k+1}(\belief) = \frac{\aprob_{\belief,\action_{k+1}}
    \,  \int_\Belief \oprob_{\state_{k+1}, \obs_{\belief_k,\belief}}\, \post_k(\belief_k) d\belief_k}
  {\int_\Belief \aprob_{\belief,\action_{k+1}}
    \,  \int_\Belief \oprob_{\state_{k+1}, \obs_{\belief_k,\belief}}\, \post_k(\belief_k) d\belief_k \, d\belief}
  \label{eq:post}
  \eeq
  Here $\obs_{\belief_k,\belief}$ is the observation such that $ \belief = \filter(\belief_k,\obs)$ where $\filter$ is the adversary's  filter (\ref{eq:belief}).  
  We call (\ref{eq:post})  the {\em optimal inverse filter} since it yields the Bayesian posterior of the adversary's  belief given our state and noisy measurements of the  adversary's actions.

\subsection*{Example: Inverse Kalman Filter} \label{sec:inversekalman}
We consider  a second special case of (\ref{eq:post}) where the  inverse filtering problem  admits a finite dimensional characterization in terms of  the Kalman filter.
Consider a  linear Gaussian state space model
\beq \label{eq:lineargaussian}
\begin{split}
\state_{k+1} &= \statem\, \state_k  + \snoise_k, \quad \state_0 \sim \belief_0 \\
\obs_k &= \obsm\, \state_k + \onoise_k 
\end{split}
\eeq
where  $\state_k \in \statespace = \reals^\statedim$ is ``our'' state with
initial density $\belief_0 \sim \normal(\hat{\state}_0,\kalmancov_0)$,
 $\obs_k \in \obspace = \reals^\obsdim$ denotes the adversary's observations,
 $\snoise_k\sim \normal(0,\snoisecov_k)$,
 $\onoise_k \sim \normal(0,
\onoisecov_k)$
and 
  $\{\snoise_k\}$,  
  $\{\onoise_k\}$ are mutually independent  i.i.d.\ processes.

 Based on observations $\obs_{1:k}$, the adversary computes the  belief  $\belief_k = \normal(\hstate_k,\kalmancov_k)$ where $\hstate_k$ is the conditional mean
  state   estimate and $\kalmancov_k$ is the covariance; these are computed via the classical Kalman filter
  equations:\footnote{For localization problems, we will use the information filter form:
    \beq  \kalmancov^{-1}_{k+1} = \kalmancov_{k+1|k}^{-1} + \obsm^\p \onoisecov^{-1} \obsm, \quad
\kg_{k+1} = \kalmancov_{k+1} \obsm^\p \onoisecov^{-1} \label{eq:info}
\eeq Similarly, the inverse Kalman filter in information form reads
\beq \enemykalmancov^{-1}_{k+1} = \enemykalmancov^{-1}_{k+1|k} +
\enemyobsm_{k+1}^\p \enemyonoisecov^{-1} \enemyobsm_{k+1},\;
\enemykg_{k+1} = \enemykalmancov_{k+1} \enemyobsm_{k+1}^\p \enemyonoisecov^{-1}. \label{eq:enemyinfo}
\eeq}
\beq
  \begin{split}
\kalmancov_{k+1|k} &=  \statem  \kalmancov_{k} \statem^\p  +  \snoisecov  \\
\Sig_{k+1} &= \obsm \kalmancov_{k+1|k} \obsm^\p + \onoisecov 
\\
{\hstate}_{k+1} &=  \statem\,  {\hstate}_k  + 
\kalmancov_{k+1|k} \obsm^{\p}  \Sig_{k+1}^{-1} 
(\obs_{k+1} - \obsm \, \statem\,  {\hstate}_k )
\\
\kalmancov_{k+1} &=
\kalmancov_{k+1|k} -  
\kalmancov_{k+1|k} \obsm^{\p}  \Sig_{k+1}^{-1} 
\obsm \kalmancov_{k+1|k} 
\end{split}
\label{eq:kalman}
\eeq
  The 
  adversary then chooses its  action as  $\eaction_k = \fun(\kalmancov_k)\,\hstate_k$ for some pre-specified function\footnote{This choice is motivated by linear quadratic Gaussian control where the action (control) is chosen as a linear function of the estimated state $\hstate_k$ weighed by the covariance matrix}
  $\fun$. We  measure the adversary's  action as
\beq \action_k = \fun(\kalmancov_k)\,\hstate_k + \anoise_k, \quad
\anoise_k \sim \text{ iid } \normal(0,\anoisecov) \label{eq:linearaction} \eeq

The Kalman covariance $\kalmancov_k$ is deterministic and fully determined by the model parameters. So  to estimate the belief
$\belief_k=\normal(\hstate_k,\kalmancov_k)$ we only need  to estimate $\hstate_k$ at each time $k$ given $a_{1:k},\state_{0:k}$.
Substituting (\ref{eq:lineargaussian})  for $\obs_{k+1}$ in (\ref{eq:kalman}), we see that
(\ref{eq:kalman}), (\ref{eq:linearaction}) constitute a linear Gaussian system with un-observed state  $\hstate_k$, observations $\action_k$,
and known exogenous  input $\state_k$:
\beq \label{eq:inversekf}
\begin{split}
  {\hstate}_{k+1} &=   (I - \kg_{k+1} \obsm) \, \statem \hstate_{k} + \kg_{k+1} \onoise_{k+1} + \kg_{k+1} \obsm \state_{k+1} \\
   \action_k &= \fun(\kalmancov_k)\,\hstate_k + \anoise_k, \quad
   \anoise_k \sim \text{ iid } \normal(0,\anoisecov) \\
  & \text{ where }  \kg_{k+1} = \kalmancov_{k+1|k} \obsm^{\p}  \Sig_{k+1}^{-1} 
\end{split}
\eeq
$
\kg_{k+1}$   is  called the Kalman gain.

To summarize,  our filtered estimate of the adversary's filtered estimate    $\hstate_k$ given measurements $a_{1:k},\state_{0:k}$ is achieved by running ``our''  Kalman filter on the linear Gaussian state space model  (\ref{eq:inversekf}), where
$\hstate_k, \kg_k, \kalmancov_k$ in (\ref{eq:inversekf}) are generated by the adversary's  
Kalman filter. Therefore, our Kalman filter uses the parameters
\beq
\begin{split}
\enemystatem_{k}  &=  (I - \kg_{k+1} \obsm)\statem, \;
\enemyinputm_k = \kg_{k+1} \obsm ,\;
  \enemyobsm_k =  \fun(\kalmancov_k) , \\
\enemysnoisecov_k  &= \kg_{k+1}\, \kg_{k+1}^\p, \;
\enemyonoisecov = \anoisecov
\end{split} \label{eq:inversekfparam}
  \eeq
 The equations of our inverse Kalman filter are:
\beq
 \begin{split}
  \enemykalmancov_{k+1|k} &=  \enemystatem_k \enemykalmancov_{k} \enemystatem_k^\p  +  \enemysnoisecov_k  \\
  \enemySig_{k+1} &= \enemyobsm_{k+1} \enemykalmancov_{k+1|k} \enemyobsm_{k+1}^\p + \enemyonoisecov  \\
   \enemystate_{k+1} &= \enemystatem_k\, \enemystate_k + 
\enemykalmancov_{k+1|k} \enemyobsm_{k+1}^{\p}  \enemySig_{k+1}^{-1} 
                        \\ & \pushright{\times \big[\action_{k+1} -   \enemyobsm_{k+1} \left(\enemystatem_{k} \hstate_k+ \enemyinputm_k \state_{k+1} \right) \big]} 
  \\
   \enemykalmancov_{k+1} &=
\enemykalmancov_{k+1|k} -  
\enemykalmancov_{k+1|k} \enemyobsm_{k+1}^{\p}  \enemySig_{k+1}^{-1} 
\enemyobsm_{k+1} \enemykalmancov_{k+1|k} 
\end{split} \label{eq:inversekfequations}
\eeq
Note $\enemystate_k$ and $\enemykalmancov_k$ denote our conditional mean estimate and covariance of the adversary's conditional mean $\hstate_k$.
The computational cost of the inverse Kalman filter is identical to the classical Kalman filter, namely $O(\statedim^2)$ computations at each time step.

  {\bf Remarks}:   
  \begin{compactenum}
   \item  As discussed in \cite{KR19},  inverse Hidden Markov model (HMM) filters and inverse particle filters can be derived. For example, the inverse HMM filter  deals with the case 
    when $\belief_k$ is computed via a Hidden Markov model (HMM) filter and the estimates of the HMM filter are observed in noise. In this case the inverse filter has a computational cost that grows exponentially with the size of the observation alphabet.
    \item In filtering it is important to determine conditions under which the filter forgets its initial condition geometrically fast (this is often called stability of the filter). In future work, it is worthwhile obtaining  minorization  conditions for  the geometric ergodicity of the inverse HMM filter, namely, conditions on our transition matrix, enemy's observation matrix and our observation of the enemy's action matrix so that the inverse filter forgets its initial condition geometrically fast.

\item A general approximation solution for (\ref{eq:post}) involves sequential Markov chain Monte-Carlo (particle filtering).
  In particle filtering, cases where it is possible to sample from the so called optimal importance function are of significant interest \cite{RAG04,CMR05}.
  In inverse filtering, our paper \cite{KR19}  shows  that the optimal importance function can be determined explicitly due to the structure of the inverse filtering problem. Specifically, in our case, the ``optimal'' importance density is  $\imp^* = \pdf(\belief_k,\obs_k| \belief_{k-1},\obs_{k-1}, \state_{k},\action_{k} )$.
Note that in our case $$\imp^*=  \pdf(\belief_k | \belief_{k-1},\obs_k)\, \pdf(\obs_k| \state_k,\action_k)
= \delta\big(\belief_k - \filter(\belief_{k-1},\obs_k)\big)\, \pdf(\obs_k|\state_k)$$ is straightforward to sample from. 
There has been a substantial amount of recent research in finite sample concentration bounds for the particle filter. We will use these results to evaluate the so called sample complexity of the resulting particle filter.
  
\end{compactenum}

\subsection{Slow Learning: Cascaded Kalman Filters and Phase-Transition}
\begin{quote} {\bf {\em How to quantify slow learning in cascaded Bayesian filters?}}
\end{quote}
In simple terms, 
inverse filtering discussed above involves cascaded optimal filters where the second filter uses the estimates generated by the first filter.
We now outline our plans to  study a remarkable property involving two cascaded Kalman filter where the first Kalman filter feeds its estimate to the second Kalman filter, and the second Kalman filter feeds its estimate to the first Kalman filter. Below we will discuss an adversarial example involving localizing a drone.

   \begin{figure}[h] \centering
      {\resizebox{13cm}{!}{
\begin{tikzpicture}[node distance = 1cm, auto]
    \node [block] (BLOCK1) {Kalman Filter 1};
    \node [block, right  of=BLOCK1,node distance=6.5cm] (BLOCK2) {Kalman Filter 2};
\draw[<-,line width=1pt] (BLOCK1.west) --  node[above]{$\obs_k= \state_0 + \onoise_k$} ++(-2.4,0);
    \draw[->,line width=1pt] (BLOCK1) -- node[above,pos=0.73]{$\action_k= \hat{\state}_k + \anoise_k$}
node[above,pos=0.1]{$\hat{\state}_k$} node[below,pos=0.1]{$\kalmancov_k$}
    (BLOCK2)  ;
    \draw[->,line width=1pt] (BLOCK2.east) -| node[above,pos=0.3] {$\hat{\hat{\state}}_k$} node[below,pos=0.3] {$\hat{\kalmancov}_k$}    ++(1.3,-1.5)  -|  (BLOCK1.south);

  \end{tikzpicture}} }

\caption{Slow Learning with Two Kalman Filters in Series. The remarkable point is that  the covariance $\kalmancov_k = O(k^{-1/3}) $ instead of $O(k^{-1})$ thereby resulting in slow learning. In our terminology, Kalman filter 2 is the inverse Kalman filter that uses  noisy estimates $\action_k$ from Kalman filter~1} \label{fig:schematicslow}
\end{figure}
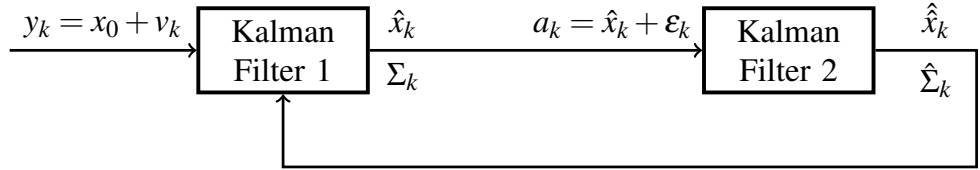

To illustrate the main idea, consider a Gaussian random variable scalar state $\state_0 \in \reals$ (instead of a random process). The first Kalman filter obtains 
 Gaussian measurements $\obs_k$ of $\state_0$. It uses $\obs_k$ together with the prior $\hat{\hat{x}}_{k-1}$ from the second Kalman filter to compute the estimate  $\hat{\state}_k$ at time $k$. The second Kalman filter
 observes  $\hat{\state}_k$ from the first Kalman filter in Gaussian noise as $\action_k = \hat{\state}_k + \anoise_k$ where $\anoise_k$ is an iid Gaussian sequence with zero mean and variance $\sigma_\anoise^2$. The second Kalman filter then  uses $\action_k$  to estimate $\state_0$. Call the estimate generated by the second Kalman filter as  $\hat{\hat{\state}}_k$. This estimate $\hat{\hat{\state}}_k$ is then fed back to the first Kalman filter. 

  Two remarkable properties of the 
 above setup are (see also \cite{Viv93,Viv17,Cha04})
\begin{compactitem}
\item The covariance decreases with time $k$ as  $\kalmancov_k = O(k^{-1/3})$ instead of $O(k^{-1}) $ in the standard Kalman filter. This depicts ``slow learning'' of the inverse Kalman filter (in Fig.\ref{fig:schematicslow}).
\item This rate of slow learning is independent of the variance $\sigma_\anoise^2$ as long as $\sigma_\anoise^2$ is strictly positive.
  Note that if $\sigma_\anoise^2=0$, then  $\kalmancov_k = O(k^{-1})$. This shows that a cascaded model of two Kalman filters where the second Kalman filter  observes the first Kalman has a phase change in behavior:
  when the estimates of the first Kalman filter are corrupted even by small amounts of noise, the convergence rate slows down drastically to $O(k^{-1/3})$.
\end{compactitem}

\subsection*{Example. Sequential Localization Game}

   Thus  far  we assumed that  $\state_0$ is known to us and our aim was to estimate the adversary's estimate. Consider now  the following  localization game.
 \begin{figure}[h] \centering
      {\resizebox{7.5cm}{!}{
\begin{tikzpicture}[node distance = 2cm, auto]
    \node [block] (BLOCK1) {Our Kalman filter};
    \node [block, right of=BLOCK1,node distance=6.5cm] (BLOCK2) {Enemy's Kalman filter};

    \node[draw=none,fill=none, above right =0.7cm and 0.75cm of BLOCK1] (drone) {\includegraphics[scale=0.4]{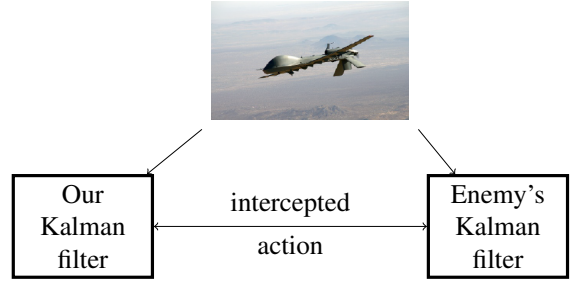}};
    
    \draw[<->] (BLOCK1) -- node[above]{intercepted} node[below]{action} (BLOCK2)  ;
 
    \draw[->] (drone.south west) -- (BLOCK1);
    \draw[->] (drone.south east) -- (BLOCK2);

  \end{tikzpicture}} }
\caption{Sequential Localization Game Schematic}
\label{fig:game}
\end{figure}
   Suppose  that
 neither us nor the adversary know the underlying state $\state_0$. For example, $\state_0$ could denote the threat level of an autonomous  drone hovering relative to a specific location as shown in Figure \ref{fig:game}. Second, there is feedback; the adversary's action affects our estimate and our action affects the adversary's estimate.
 The setup is naturally formulated in game-theoretic terms.
 The  adversary and us play a sequential game  to localize  (estimate)~$\state_0$.

 \begin{compactenum}
 \item At odd time instants $k=1,3,\ldots$, the  adversary takes active  measurements of the target  and makes decisions based on its estimate.  We  eavesdrop on the adversary's decisions and apply Bayes rule to estimate (localize) the target's state $\state_0$. The details are as follows: \\ (i)  The
 adversary  observes our action $\laction_{k-1} = \hstate_{k-1}$ in noise as  $\action_{k-1}= \hstate_{k-1}+ \anoise_{k-1}$ where $\anoise_{k-1} \sim \normal(0,\anoisecov)$. The adversary assumes that our state estimate is $\action_{k-1}$. 
 \\ (ii) The adversary  takes a direct measurement $\obs_k$ of $\state_0$  and uses
 its Kalman filter  (\ref{eq:kalman}) to 
 update its  state  estimate:
$$\hstate_k = (1 - \kg_{k}) \action_{k-1} + \kg_{k} \obs_k, $$
 where $\kg_k $ is the associated covariance update.\\
(iii) 
The adversary then 
takes action $\laction_k= \hstate_k$. This action  minimizes the mean square error of the estimate. \\
(iv) 
We eavesdrop on (observe)  the adversary's action in noise as $\action_k=\hstate_k + \anoise_k$. Assume $\anoise_k \sim \normal(0,\anoisecov)$ iid. So
$$ \action_{k} = (1 - \kg_{k}) \action_{k-1} + \kg_{k} \onoise_{k} +
\kg_{k} \state_0 + \anoise_{k} $$
Since  $\action_{k-1} = \laction_{k-1}+ \anoise_{k-1}$, we can define an observation $\eobs_k$ that is informationally equivalent to $\action_k$ as
\beq \eobs_k \ole  \frac{\action_k }{\kg_k} - \frac{1-\kg_k}{\kg_k}\,\laction_{k-1} 
= \frac{1-\kg_k}{\kg_k}\,\anoise_{k-1}+ \onoise_k + \frac{\anoise_k}{\kg_k} + \state_0
\label{eq:eobs}
\eeq
We know  $\eobs_k$; $\laction_{k-1}$ was our action at time $k-1$, and $\action_k$ is our measurement of the adversary's action at time $k$.

To summarize, at odd time instants,
 $\eobs_k $ specified in (\ref{eq:eobs})  is our effective observation of the unknown state $\state_0$ purely based on sensing the adversary's actions.
\item At even time instants $k=2,4,\ldots,$  we take active measurements $\obs_k$ of the target's state $\state_0$. Then we update our estimate of our target using the Kalman filter.
\end{compactenum}
The above setup constitutes a social learning sequential game \cite{MMST18}. Since the decisions are made myopically each time (to minimize the mean square  error), the strategy profile at each time constitutes a Nash equilibrium. More importantly \cite[Theorem 5]{MMST18}, the asymptotic behavior (in time) is captured by a social learning equilibrium (SLE).

\begin{theorem}[\cite{KR19}] \label{thm:gamecov}
  Consider the sequential game formulation for localizing the random variable state $\state_0$. Then for large $k$,
  the covariance of our estimate of the state $\state_0$ is
  $\kalmancov_k = 2 k^{-1}$ (which is twice the covariance for classical localization).
\end{theorem}

{\em Remark}. Theorem~\ref{thm:gamecov} has two interesting consequences. \begin{compactenum}
  \item The asymptotic covariance is independent of $\anoisecov$ (i.e.  SNR of  our observation of the adversary's action)
as long as $\anoisecov > 0$. 
If $\anoisecov = 0$, there is a phase change and the covariance $\kalmancov_k = k^{-1}$
as in the standard Kalman filter.
\item 
  In the special case where  only (\ref{eq:eobs}) is observed at each time
(and the $\anoise_{k-1}$ term is omitted),
  the
 formulation reduces to the Bayesian social learning problem 
  of \cite[Chapter 3]{Cha04}. In that case, \cite[Proposition 3.1]{Cha04} shows that  $\kalmancov_k = O(k^{-1/3})$.  The remarkable property is that this rate of slow learning is independent of the variance $\sigma_\anoise^2$ as long as $\sigma_\anoise^2$ is strictly positive \cite{Viv93,Viv17}. As discussed in \cite{Cha04}, this result implies that models where one assumes that the enemy's action is observed perfectly are not robust; any noise in the observation process drastically reduces the rate of convergence of the estimator.
\end{compactenum}
To summarize, Theorem \ref{thm:gamecov} confirms the intuition  that sequentially learning the state indirectly from the adversary's actions (and the adversary learning the state from our actions) slows down the convergence of the localization estimator.

The above setup constitutes a social learning sequential game \cite{MMST18}. Since the decisions are made myopically each time (to minimize the mean square  error), the strategy profile at each time constitutes a Nash equilibrium. More importantly \cite[Theorem 5]{MMST18}, the asymptotic behavior (in time) is captured by a social learning equilibrium (SLE).

\subsection{Estimating the Enemy's Sensor Gain}
Thus far we have 
 discussed Bayesian estimation of  the adversary's belief state.  We now discuss how to  estimate the enemy's sensor observation kernel  $\oprob$ in (\ref{eq:model}) which  quantifies the accuracy of  the adversary's sensors.
We assume that  $\oprob$   is parametrized by an $\modeldim$-dimensional vector $\model \in \Model$ where
$\Model$ is a compact subset of $\reals^\modeldim$. Denote the parametrized
observation kernel as $\oprob^\model$.
Assume  that both us and the adversary know $\tp$ (state transition kernel\footnote{As mentioned earlier, otherwise the adversary estimates $\tp$ as $\hat{\tp}$ and we need to estimate the adversary's estimate as
  $\hat{\hat{\tp}}$. This makes the estimation task substantially more complex.}) and $\aprob$ (probabilistic map from adversary's belief to its action).
Then given our state sequence $\state_{0:\horizon}$  and adversary's action sequence $\action_{1:\horizon}$, our aim is to compute the maximum likelihood estimate (MLE) of $\model$. That is, with $\lik(\model)$ denoting the log likelihood, the aim is to compute
$$ \mle = \argmax_{\model \in \Model} \lik(\model), \quad \lik(\model)= \log \pdf(\state_{0:\horizon},\action_{1:\horizon} | \model) .$$
The likelihood can be evaluated from
 the un-normalized  inverse filtering recursion (\ref{eq:post}) 
 \beq   \lik(\model) = \log \int_\Belief \unpostm_\horizon(\belief) d\belief,
 \quad 
\unpostm_{k+1}(\belief) = \aprob_{\belief,\action_{k+1}}
    \,  \int_\Belief \oprob^\model_{\state_{k+1}, \obs^\model_{\belief_k,\belief}}\, \unpostm_k(\belief_k) d\belief_k
\label{eq:unpost}
\eeq

Given  (\ref{eq:unpost}), a local stationary point of the likelihood
can be computed using  a general purpose numerical optimization algorithm.

{\em Remark. EM Algorithm is not useful}: The Expectation Maximization (EM) algorithm is a popular numerical algorithm for computing the MLE especially when the Maximization (M) step can be computed in closed form. Unfortunately, for estimating the adversary's observation model,  due to the time evolving dynamics in the inverse Kalman filter, the EM algorithm is not useful
since the M-step involves a non-convex optimization that cannot be done in closed form. There is no obvious way of choosing the latent variables to avoid this. 

\section*{Example. Estimating Adversary's Gain in Linear Gaussian case}

Consider the setup  in Sec.\ref{sec:inversekalman} where our dynamics are linear Gaussian and the adversary observes our state linearly in Gaussian noise (\ref{eq:lineargaussian}). The adversary
estimates our state using a Kalman filter, and we estimate the adversary's estimate using the inverse Kalman filter (\ref{eq:inversekf}).
Using  (\ref{eq:inversekf}), (\ref{eq:inversekfparam}),  the log likelihood  for the adversary's observation  gain matrix $\param=\hobsm$  based on our measurements is
\begin{align}
  \lik(\model) &= \text{const} - \frac{1}{2} \sum_{k=1}^\finaltime \log | \enemySig^\param_k| - \frac{1}{2} \sum_{k=1}^\finaltime
  \innovations_k^\p\, (\enemySig_k^\param)^{-1} \,\innovations_k\nonumber \\
  \innovations_k &= \action_k - \enemyobsm_k^\param\, \enemystatem_{k-1}^\param \enemystate_{k-1} - \enemyinputm_{k-1}^\param \state_{k-1} \label{eq:inversekflik}
\end{align}
where $\innovations_k$ are the innovations of the inverse Kalman filter (\ref{eq:inversekfequations}).
In (\ref{eq:inversekflik}), our state $\state_{k-1}$ is known to us and therefore is a known exogenous input. Also note from (\ref{eq:inversekfparam}) that $\enemystatem_{k}, \enemyinputm_k$
are explicit functions of $\obsm$, while $\enemyobsm_k$ and $\enemysnoisecov_k$ depend on $\obsm$ via the adversary's Kalman filter.

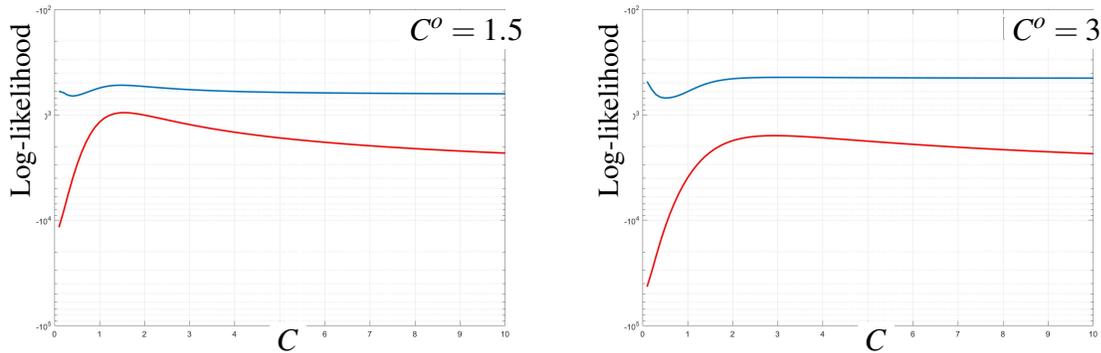
\begin{figure}[h]
   \begin{overpic}[scale=0.11,unit=1mm]{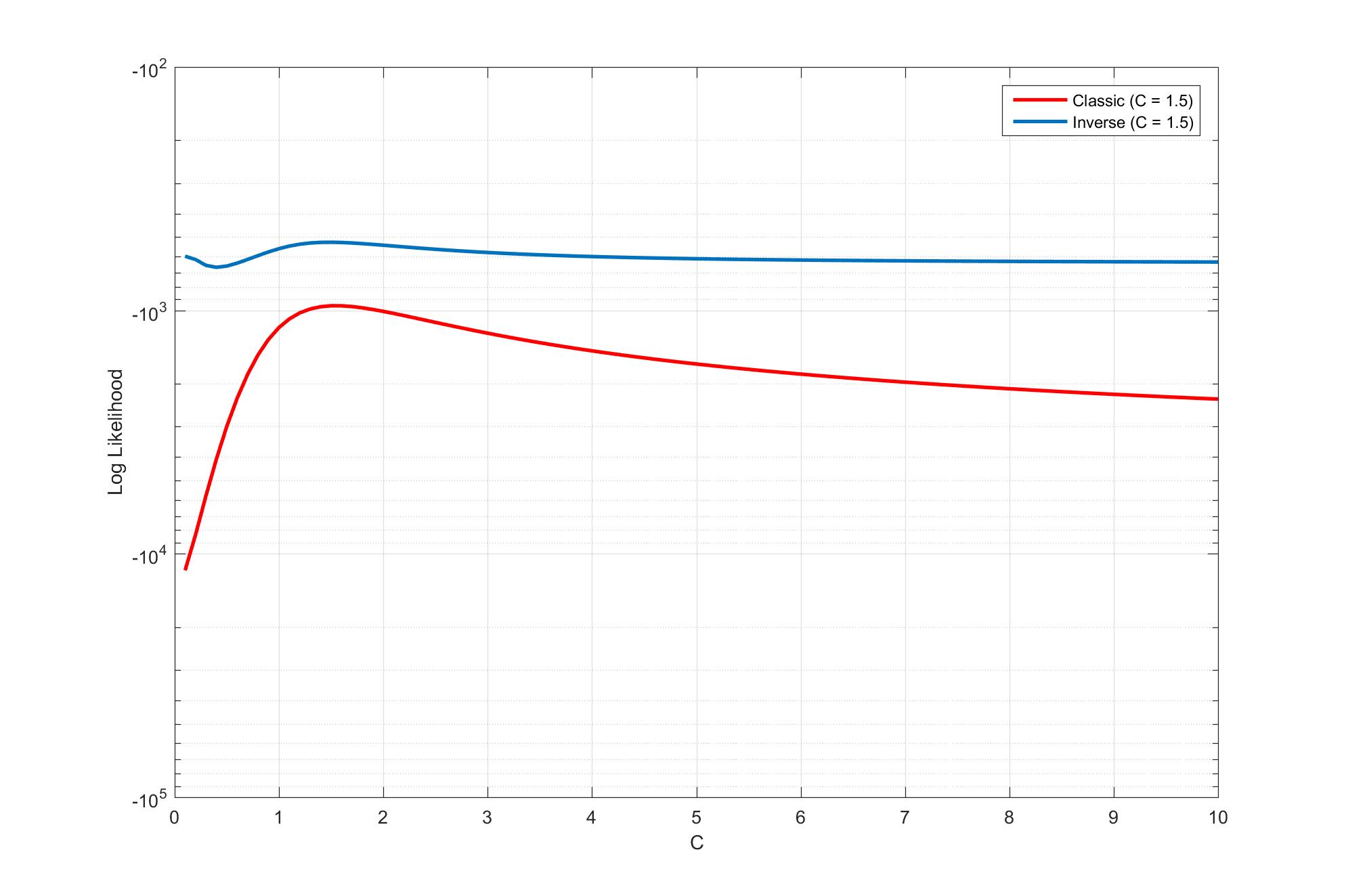}
    \put(50.15,2.9){\colorbox{white}{$\hobsm$}}
    \put(4,28){\rotatebox{90}{\colorbox{white}{\small Log-likelihood}}}
     \put(73,56.5){\colorbox{white}{$\obsmt=1.5$}}
  \end{overpic}
  \begin{overpic}[scale=0.11,unit=1mm]{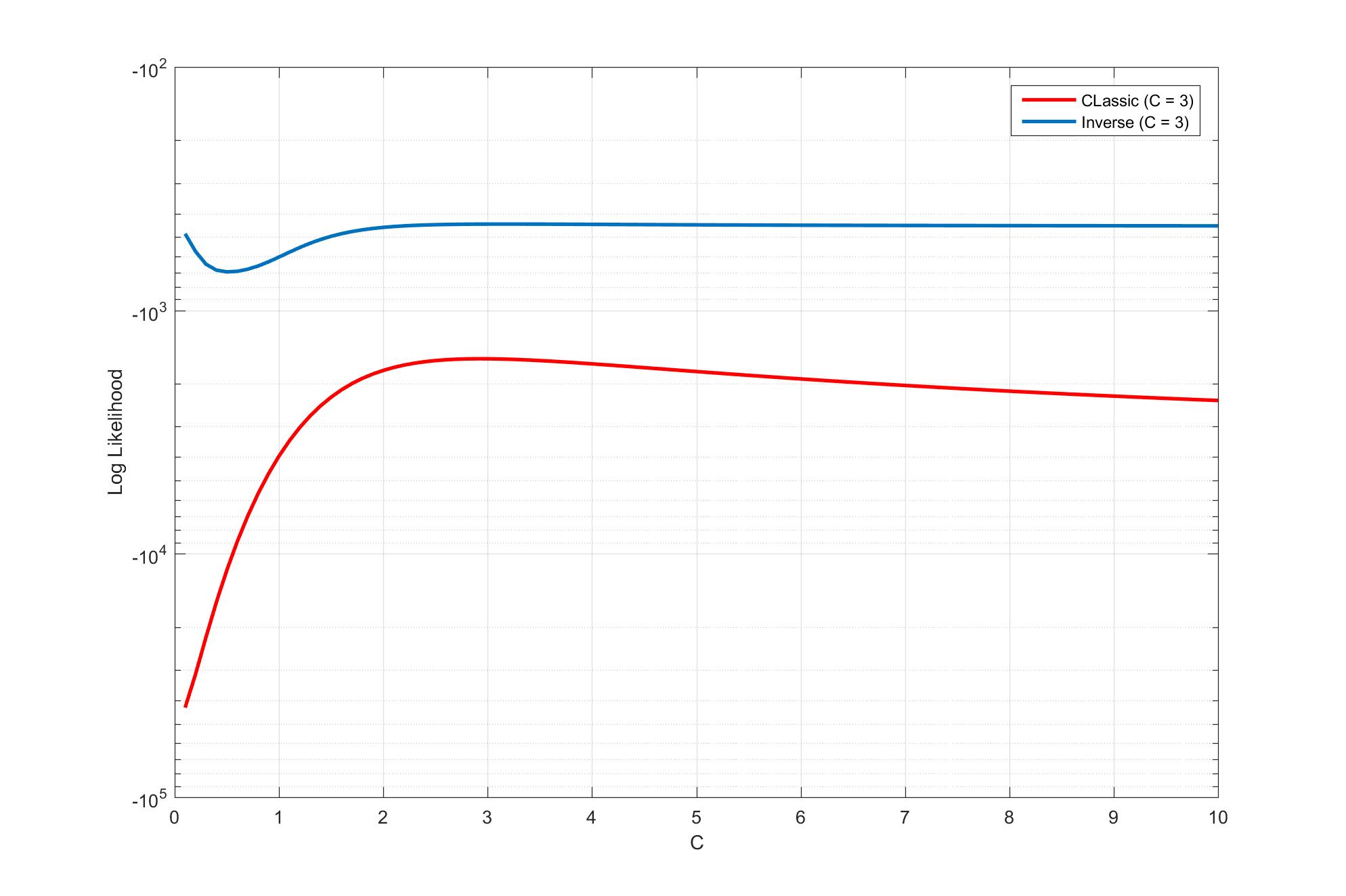}
    \put(50.15,2.9){\colorbox{white}{$\hobsm$}}
    \put(4,28){\rotatebox{90}{\colorbox{white}{\small Log-likelihood}}}
     \put(75.5,56.5){\colorbox{white}{$\obsmt=3$}}
  \end{overpic}
\caption{Log-Likelihood as a function of enemy's gain $\hobsm\in (0,10]$ when true value is $\obsmt$. The red curves denote the log-likelihood of $\hobsm$ given the enemy's measurements. The blue curves denotes the log-likelihood of $\hobsm$ using the inverse Kalman filter given our observations of the enemy's action. The plots show that it is more difficult to compute the MLE for the inverse filtering problem due to the almost flat likelihood (blue curves) compared to red curves.}
\label{fig:kf}
\end{figure}

Using (\ref{eq:unpost}), the log likelihood  for the adversary's observation  gain matrix $\param=\hobsm$  can be evaluated
To provide insight,  Figure \ref{fig:kf} displays  the log-likelihood versus adversary's gain matrix $\hobsm$  in the scalar case for 1000 data points.
 The four sub-figures correspond to true values of $\obsmt = 1.5,3$ respectively.

Each sub-figure in  Figure \ref{fig:kf} has two plots. The plot in red is  the  log-likelihood of $\hat{\obsm} \in (0,10]$ evaluated based on the adversary's observations using the  standard  Kalman filter. (This is the classical log-likelihood of the observation gain of a Gaussian state space model.)  The plot in blue is the log-likelihood of $\hobsm\in (0,10]$ computed using  our measurements of the adversary's action using the  inverse Kalman filter (where the adversary first estimates our state using a Kalman filter) - we call this the inverse case.

 Figure \ref{fig:kf} shows  that the log likelihood in the inverse case (blue plots) has a less pronounced maximum compared to the standard case (red plots). Therefore, numerical algorithms for computing the MLE of the  enemy's gain using our observations of the adversary's actions (via the inverse Kalman filter) will converge much more slowly than the classical MLE (based on  the adversary's observations). This is intuitive  since  our estimate of the adversary's parameter is based on the adversary's estimate of our state and so has more noise.

{\bf Cramer Rao (CR) bounds}. Is it instructive to compare the   CR bounds for MLE of $\obsm$ for the classic model versus that of the inverse Kalman filter model. Table \ref{tab:cr} displays the CR bounds (reciprocal of expected Fisher information) for the four examples considered above evaluated using via the algorithm in  \cite{CS96}. 
It shows that the covariance lower bound for the inverse case is substantially higher than that for the classic case. This is
consistent with the intuition that estimating the adversary's parameter based on its actions (which is based on its estimate of us) is more difficult than directly estimating $\obsm$ in a classical state space model.

\begin{table} \centering
\begin{tabular}{|c|c|c|}
  \hline
  $\obsmt$ & Classic & Inverse \\
  \hline
  0.5 & $0.24 \times 10^{-3} $ & $5.3 \times 10^{-3}$ \\
  1.5 & $1.2 \times 10^{-3}$ &   $37 \times 10^{-3}$ \\
  2    & $2.1 \times 10^{-3} $ &  $70 \times 10^{-3}$ \\
  3  & $ 4.6 \times 10^{-3}$ & $ 336 \times 10^{-3}$\\
  \hline                       
\end{tabular}
\caption{Comparison of Cramer Rao bounds for $\obsm$ - classical  model vs inverse Kalman filter model}
\label{tab:cr}
\end{table}

{\em Remarks}:
\begin{compactenum} \item
  {\em Consistency of MLE}. The above example shows that the likelihood surface
of  $ \lik(\model)= \log \pdf(\state_{0:\horizon},\action_{1:\horizon} | \model)$ is flat and hence compute the MLE numerically can be difficult.  Even in the case when we observe the enemy's actions perfectly, our NIPS paper \cite{MRKW17} shows that non-trivial  observability conditions need to be imposed on the system parameters.
  
For the linear Gaussian case where we observe the enemy's Kalman filter in noise,
consistency  of the MLE for the for the adversary's gain matrix $\obsm$, the strong
consistency can be shown fairly straightforwardly. Specifically, if we 
assume that state matrix $\statem$ is stable, and the state space model 
is an identifiable minimal realization, then the enemy's Kalman filter variables 
 converge to steady state values geometrically fast in  $k$  \cite{AM79} implying that asymptotically the inverse Kalman filter system  is stable linear time invariant. Then , the  MLE $\mle$  for the adversary's observation matrix $\obsm$   is unique and a strongly consistent estimator \cite{Cai88}.

\item \textit{Estimating the enemy's estimate of our dynamics (transition matrix)}
  It is of interest to develop algorithms  to estimate the enemy's estimate of our
  transition matrix and analyze their performance. The adversary estimates $\tp$ as $\hat{\tp}$ and we need to estimate the adversary's estimate as
  $\hat{\hat{\tp}}$. In future work we will examine conditions under which the MLE is identifiable  and consistent. 
\end{compactenum}

\newpage
\section{Inverse Cognitive Radar -- Revealed Preferences and Inverse Reinforcement Learning}

The previous section was concerned with estimating the enemy's posterior belief and sensor accuracy.  
This section discusses detecting utility maximization behavior and estimating the enemy's utility function
in the context of cognitive radars.

Cognitive radars  \cite{Hay06} use  the perception-action cycle of cognition to  sense the environment, learn from it relevant information about the target and the background, then adapt the radar sensor to optimally satisfy the needs of their mission. A
crucial element of a cognitive radar  is optimal adaptivity: based on its tracked estimates, the radar   adaptively optimizes 
the waveform, aperture, dwell time and revisit rate. In other words, a cognitive radar is a constrained utility maximizer.

This section   is motivated by the  next logical step, namely, \textit{inverse cognitive radar}.
 The enemy's cognitive radar
observes our  state in noise; it  uses a Bayesian estimator (target tracking algorithm)  to update its posterior
distribution of our state and then chooses an action based on this
posterior.
From the intercepted emissions of an enemy's radar: 
\begin{compactenum}
  \item[\textbf{A:}]
    Are the enemy sensor's actions consistent with optimizing a monotone utility function (i.e., is the cognitive sensor behavior rational in an economics sense)? If so how can we estimate a utility function of the enemy's  cognitive sensor that is consistent with its actions?
    \item[\textbf{B:}] How to construct a statistical detection test for utility maximization when we observe the enemy sensor's actions in noise?
\item[\textbf{C:}]  How can we optimally probe the enemy's sensor by choosing our state   to minimize the Type 2  error of detecting if the radar is deploying an economic rational strategy, subject to a constraint on the Type 1 detection error?
  ``Our'' state can be viewed as a probe signal which causes the enemy's sensor  to act.
  \item[\textbf{D:}] How to detect utility maximization of a Bayesian adversary that is rationally inattentive? Rational inattention (from behavioral economics) introduces an information theoretic cost to the accuracy of an observation.
\end{compactenum}
The main synthesis/analysis framework we will use   is  that of revealed preferences \cite{Var12,FM09,Die12} from microeconomics which
aims to determine preferences by observing choices. The results presented below are developed in detail in our recent paper \cite{KAEM20}.

    \begin{figure}[h] \centering
            {\resizebox{8cm}{!}{
\begin{tikzpicture}[node distance = 1cm, auto]
    \node [blocka] (BLOCK1) {Sensor};
    \node [blocka, below of=BLOCK1,right of=BLOCK1,node distance=1.5cm] (BLOCK2) {Optimal Decision \\ Maker};
    \node [blocka, below of=BLOCK1,left of=BLOCK1,node distance=1.5cm] (BLOCK3) {Bayesian Tracker};

    \draw[<-] (BLOCK1) -| node[left,pos=0.8]{$\response_\dtime$}  (BLOCK2)  ;
    \draw[->] (BLOCK1.west) -|   node[left,pos=0.8]{$\obs_k$} (BLOCK3);

    \draw[->](BLOCK3) --  node[above]{$\belief_k$} (BLOCK2);

    \node[draw=none,fill=none] at (4.5,-1.5) (drone) {\includegraphics[bb=0 0 0 0,scale=0.4]{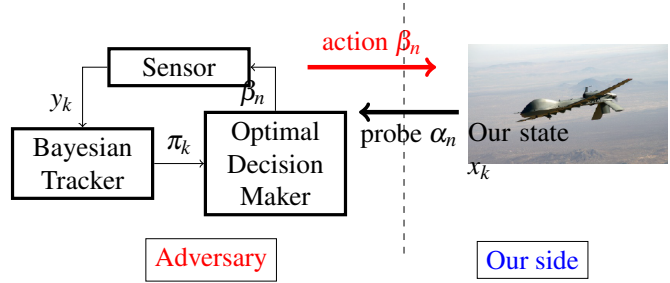}};
     \node[text width=2cm] at (5.5,-1.3) {{Our state $\state_k$}};

    \draw[->,color=red,line width=2pt] (2,0)   -- node[above]{action $\response_\dtime$}(4,0);
    \draw[->,line width=2pt] ([yshift=0.8cm]drone.west)   --   node[below]{probe $\probe_\dtime$} (2.8,-0.7);
    \node[draw] at (5.5,-3.0) {{\color{blue}Our side}};
    \node[draw] at (0.5,-3.0) {{\color{red}Adversary}};
    \draw [dashed] (3.5,1) -- (3.5,-3);
   \end{tikzpicture}} }

\caption{Schematic of Adversarial Inference Problem. Our side is a drone/UAV or electromagnetic signal that probes the  enemy's cognitive radar system. $k$ denotes a fast time scale and $n$ denotes a slow time scale. Our state $\state_k$, parameterized by $\probe_\dtime$ (purposeful acceleration maneuvers), probes the adversary radar. Based on the noisy observation $\obs_k$ of our state, the enemy radar  responds with action~$\response_\dtime$. Our aim is to determine if the enemy radar is economic rational, i.e.,  is its response  $\beta_\dtime$  generated by constrained optimizing a utility function?}
\label{fig:schematica}
\end{figure}

{\bf Classical Inverse Reinforcement Learning (IRL) vs Revealed Preferences}. Since we focus on revealed preferences,  it is worthwhile comparing it with inverse reinforcement learning.

\begin{itemize}
\item Inverse reinforcement learning  (IRL)  aims to estimate the cost  vector $c$ (which depends only on the state)
  of an infinite horizon Markov decision process (MDP) by observing the decisions and knowing the controlled transition probabilities \cite{NR00}. The idea in \cite{NR00} is  simple: If the decisions $\action$ are chosen from optimal policy $\mu^*$, then the infinite horizon cumulative cost satisfies
  $L(\mu^*, c) \leq L(\mu, c) $ for any policy $\mu$. Also $L(\mu,c)$ is a linear function of cost vector $c$, i.e., $L(\mu,c) = H(\mu) c$ where matrix $H(\mu)$  depends on the transition matrix and discount factor.  Since there are only a finite number of possible policies $\mu$ (as the state and action space are finite), clearly the cost vector $c$ satisfies the finite number of  linear inequalities:
 \beq  [ H(\mu^*) - H(\mu)]  c \leq  0 ,  \quad \mu = 1,2,\ldots,\actiondim   \label{eq:irl} \eeq
Note also that in  IRL, the
   existence of a utility/cost function is assumed implicitly.
   \item The revealed preference approach that we consider
addresses a deeper and more fundamental question: {\em does a utility function exist that rationalizes the
given data with  budget  constraints? If so, estimate this utility function}. Indeed IRL  (\ref{eq:irl}) is a trivial case of the revealed preference framework. It is surprising that despite powerful revealed preference methods being developed since the 1960s (e.g.\ Afriat's theorem), and numerous advances subsequently,  the machine learning community is completely oblivious to this. Finally we note that the first result in IRL was developed by Kalman \cite{Kal64} in the 1964 under the terminology ``inverse optimal control''.
\end{itemize}

\subsection{Background. Revealed Preferences and Afriat's Theorem} Nonparametric detection of utility maximization behavior is the central theme in the area of revealed preferences in microeconomics.
We briefly outline a key result.
\begin{definition}[\cite{Afr67,Afr87}]\label{eq:utility_maximizer}
A system is  a {\em utility maximizer} if for every probe $\probe_\dtime\in \reals_+^\probedim$, the  response $\response_\dtime \in \reals^\probedim$ satisfies
\begin{equation}
  \response_\dtime\in \argmax_{\probe_\dtime^\p \response \leq 1}\utility(\response) 
\label{eq:utilitymaximization}
\end{equation}
where $\utility(\response)$ is a {\em monotone} utility function.
 \end{definition}

In economics, $\probe_\dtime$ is the price vector and  $\response_\dtime$ the consumption vector. Then $\probe_\dtime^\p \response \leq 1$ is a natural budget constraint\footnote{The budget constraint $\probe_\dtime^\p \response \leq 1$ is without loss of generality, and can be replaced by $\probe_\dtime^\p \response \leq c$ for any positive constant $c$. A more general nonlinear budget incorporating spectral constraints will be discussed  below.}  for a consumer with 1 dollar. Given a dataset of price and consumption vectors, the aim is to determine
   if the consumer is a utility maximizer (rational) in the sense of (\ref{eq:utilitymaximization}).  
 
 The key result  is the 
following remarkable theorem  
due to Afriat~\cite{Die12,Afr87,Afr67,Var12,Var83}

\begin{theorem}[Afriat's Theorem~\cite{Afr67}] Given a data set
$
 \dataset=\{(\probe_\dtime,\response_\dtime), \dtime\in \{1,2,\dots,\horizon\}\}$,
 the following statements are equivalent:
	\begin{compactenum}
	\item The system is a utility maximizer and there exists a monotonically increasing,\footnote{We use monotone and local non-satiation interchangeably. Afriat's theorem was originally stated for a non-satiated utility function which is slight generalization of monotone.} continuous,  and concave utility function by satisfies (\ref{eq:utilitymaximization}). 
		\item For $u_t$ and $\lagrange_t>0$ the following set of inequalities has a feasible solution:
			\begin{equation}
				u_\sindx-u_\tindx-\lambda_\tindx \probe_\tindx^\p (\response_\sindx-\response_\tindx) \leq 0 \; \forall \tindx,\sindx\in\{1,2,\dots,\horizon\}.\
				\label{eqn:AfriatFeasibilityTest}
			\end{equation}
		\item Explicit  monotone and concave utility
                  functions\footnote{As pointed out in Varian's influential paper \cite{Var82}, a remarkable feature of Afriat's theorem is that if the dataset can be rationalized by a monotone utility function, then it can be rationalized
by a continuous, concave, monotonic utility function. Put another way,   continuity and  concavity cannot be refuted with  a finite datasaset.}  that rationalize the dataset by satisfying (\ref{eq:utilitymaximization}) are given by:
			\begin{equation}
				\utility(\response) = \underset{\tindx\in \{1,2,\dots,\horizon\}}{\operatorname{min}}\{u_\tindx+\lambda_\tindx \probe_\tindx^\p(\response-\response_\tindx)\}
				\label{eqn:estutility}
                              \end{equation}
                              where $u_\tindx$ and $\lambda_\tindx$ satisfy the linear  inequalities (\ref{eqn:AfriatFeasibilityTest}).
                      \item The data set $\mathcal{D}$ satisfies the Generalized Axiom of Revealed Preference (GARP) also called cyclic consistency, namely for any $\tindx \leq \horizon$,
                       $\probe_t^\p \response_t \geq \probe_t^\p \response_{t+1} \quad \forall t\leq k-1 \implies \probe_k^\p  \response_k \leq \probe_k^\p  \response_{1}$.
	\end{compactenum}
\label{thm:AfriatTheorem}
\end{theorem}

Afriat's theorem tests for economics-based rationality; its  remarkable property is that it gives a {\em necessary and sufficient condition} for a system  to be a utility maximizer based on the system's input-output response. 
The feasibility of the set of inequalities (\ref{eqn:AfriatFeasibilityTest}) can be checked using a linear programming solver; alternatively GARP   can be checked  using Warshall's algorithm with $O(\horizon^3)$ computations~\cite{Var06}~\cite{Var82}. 

The recovered utility using~\eqref{eqn:estutility}  is not unique; indeed  any positive monotone increasing transformation of~\eqref{eqn:estutility} also satisfies Afriat's Theorem; that is, the utility function constructed is ordinal. This is the reason why the budget constraint $\probe_\dtime^\p \response \leq 1$ is without generality; it can be scaled by an arbitrary positive  constant and  Theorem \ref{thm:AfriatTheorem} still holds.  In signal processing terminology, Afriat's Theorem can be viewed as set-valued system identification of an \emph{argmax} system; set-valued since (\ref{eqn:estutility}) yields a set of utility functions that rationalize the finite dataset $\dataset$.

\subsection{Framework. Cognitive Radar as a Utility Maximizer}
\textit{Our working assumption is that a cognitive sensor satisfies economics-based rationality}; that is, a cognitive sensor is  a utility maximizer in the sense  of (\ref{eqn:estutility}) with  a possibly nonlinear  budget constraint. The main question then is:

\begin{quote} \textit{How to use revealed preferences as a systematic method  to detect cognitive radars? }
\end{quote}

\begin{figure}[h] \centering
\resizebox{11cm}{!}{ 
\begin{tikzpicture}[node distance = 4.5cm, auto]
  \node [block2] (BLOCK1) {``Our'' State  with parameter $\probe_\dtime$ };
    \node [block2, below of=BLOCK1,node distance=1.5cm] (BLOCK2) {Radar Controller};
    \node[block2, right of=BLOCK2] (grad) {Tracking Algorithm};
 
    \path [line] (BLOCK1.east) --++ (5.5cm,0cm) node[pos=0.15,above]{$\state_k$}  |-   node[pos=0.35,left]{$\obs_k$}    (grad);
    
    \path [line] (BLOCK2.west) --++ (-2cm,0cm)  node[red, pos=0.4,below]{$\response_\dtime$}  |-
    (BLOCK1.west);
     \path [line] (grad) --   node[pos=0.15,below]{$\belief_k$}    node[red, pos=0.8,above]{$\probe_\dtime$}  (BLOCK2);
     \node[draw, thick, dotted, inner sep=3ex, yshift=-1ex,
     fit=(BLOCK2) (grad)] (box) {};
     \node[fill=white, inner xsep=1ex] at (box.south) {Enemy's Cognitive Radar};
   \end{tikzpicture}
 }
 \caption{Interaction of our dynamics with adversary's cognitive radar. The cognitive radar comprises a Bayesian tracker which computes posterior $\belief_k$, and a radar controller that chooses action $\response_k$. Based on observing the time series $(\probe_\dtime, \response_\dtime), \dtime = 1,\ldots,\horizon$, our goal is to determine if the radar controller is a utility optimizer in the sense of (\ref{eq:utilitymaximization}). }
 \label{fig:setup}
\end{figure}
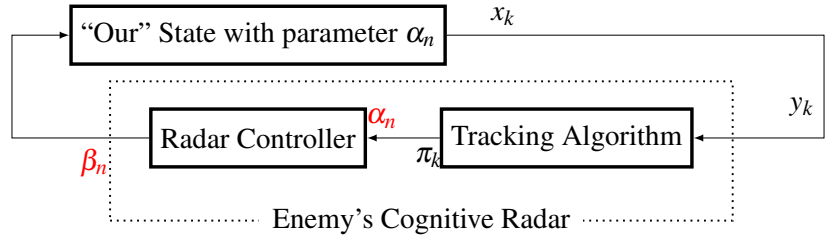

We briefly discuss the key ideas in a simplified setting.
Let $k=1,2,\ldots$ denote discrete time (fast time scale) and $\dtime=1,2,\ldots$ denote epoch (slow time scale). The framework is as follows, see Fig.\ref{fig:setup}:
\begin{compactenum}
\item Our state   $\state_k$ is parametrized by our probe signal $\probe_\dtime$. In a classical linear Gaussian state space model used in target tracking \cite{BLK08}, our probe $\probe_\dtime$ parametrizes  the state noise covariance $\snoisecov(\probe_\dtime)$  which models acceleration maneuvers of our drone.
  \item The enemy's radar controller chooses an action $\response_\dtime$
    (e.g.\ radar waveform) by optimizing a utility function $U(\response)$ subject to a possibly nonlinear constraint wrt $(\probe_\dtime,\response)$.
In a  classical target tracking model, $\response_\dtime$ parametrizes  the observation noise covariance $\onoisecov(\response_\dtime)$. From a practical point of view, when a radar adapts  its waveform function\footnote{For details of ambiguity functions and waveform design we refer to \cite{KE94}. It is here that important practicalities regarding actual radar operation become important. As mentioned earlier, our collaborator Dr. Rangaswamy at AFRL will provide us with a sophisticated test bed to emulate radar waveforms in a realistic environment.} $\response_\dtime$, in effect it adapts the measurement noise covariance $\onoisecov(\response_\dtime)$. 

\item Given the above linear Gaussian state space model with state noise covariance matrix
$\snoisecov(\probe_\dtime)$ and observation noise covariance matrix $\onoisecov(\response_\dtime)$, 
the enemy's cognitive  radar records measurements $\obs_k$ of our   kinematic state $\state_k$. The enemy uses a Kalman filter tracking algorithm to  compute the filtered posterior density $\belief_k = \normal(\hstate_k,\kalmancov_k)$ where $\hstate_k$ is the conditional mean
 state   estimate and $\kalmancov_k$ is the covariance of the estimate.
\end{compactenum}

It is well known that under the assumption that the linear Gaussian state space model parameters satisfy detectability and stabilizability \cite{AM79},
the asymptotic covariance $\kalmancov_{k+1k|k}$ as
$k\rightarrow \infty$ is the unique non-negative definite solution of  the algebraic Riccati equation (ARE):
\beq
     \ARE(\probe,\response,\kalmancov) \ole      - \kalmancov + \statem  \big(\kalmancov -  
\kalmancov \obsm^{\p}  \left[ \obsm \kalmancov \obsm^\p + \onoisecov(\response) \right]^{-1}
\obsm \kalmancov \big)  \statem^\p  +  \snoisecov(\probe) = 0
\label{eq:are}
\eeq
where $\probe_\dtime$ and $\response_\dtime$ are the probe and response signals
of the radar at epoch $\dtime$. 
Since $\kalmancov$ is parametrized by $\probe,\response$, we 
write the solution of the ARE at epoch $\dtime$ as  $\kalmancov_\dtime^*(\probe,\response)$. We will use the ARE (\ref{eq:are}) to describe our research plans in spectral revealed preferences below.

\subsection*{Example. Testing for Cognitive Radar: Spectral Revealed Preferences with  Linear Budget} \label{sec:linear}
We now show that Afriat's theorem (Theorem \ref{thm:AfriatTheorem}) can be used to determine if a radar is cognitive. The assumption here is that the utility function $\utility(\response)$ maximized by the radar is a monotone function (unknown to us) of the predicted covariance of the target.
Our main task is to formulate and justify a linear budget constraint $\probe_\dtime^\p \response \leq 1$ in Afriat's theorem.

Specifically,
suppose
\begin{compactenum}
\item Our probe  $\probe_\dtime$ that characterizes our maneuvers, is the vector of eigenvalues of the positive definite matrix
  $\snoisecov$
\item The radar response $\response_\dtime$ is the vector of eigenvalues of the positive definite matrix~$\onoisecov^{-1}$.
\end{compactenum}
Then the
cognitive radar chooses  its  waveform parameter $\response_\dtime$  at each slow time epoch $\dtime$  to maximize a utility $\utility(\cdot)$:  \beq \response_\dtime\in \argmax_{\probe_\dtime^\p \response \leq 1}\utility(\response) \label{eq:radaropt} \eeq
where $\utility$ is a monotone increasing function of $\response$.

Then Afriat's theorem   (Theorem \ref{thm:AfriatTheorem}) can be used to detect utility maximization and construct a utility function that rationalizes the response of the radar.
Recall that  the 1 in the right hand side of the budget $\probe_\dtime^\p \response \leq 1 $ can be replaced by any non-negative constant.

It only remains to justify the linear budget constraint $\probe_\dtime^\p \response \leq 1$
in (\ref{eq:radaropt}).  The $i$-th component of $\probe$, denoted as $\probe(i)$,  is the incentive for considering  the $i$-th mode of the target;  $\probe(i)$ is proportional to the signal power. The $i$-th component of $\response$ is the amount of resources (energy)  devoted by the radar to this $i$-th mode; a higher $\response(i)$ (more resources) results in a smaller  measurement noise covariance, resulting in higher accuracy of measurement by the radar.
So $\probe_\dtime^\p \response$ measures  the signal to noise ratio (SNR) and the budget constraint $\probe_\dtime^\p \response \leq 1$ is a bound on the SNR.
A rational  radar  maximizes a utility  $\utility(\response)$ that is monotone increasing in the accuracy (inverse of noise power) $\response$.  However, the radar  has limited resources and can only expend sufficient resources to ensure that the precision (inverse  covariance) of all modes is at most  some pre-specified precision  $\bound ^{-1}$ at each epoch $\dtime$.
We can then justify the linear budget constraint as follows:

\begin{lemma} \label{lem:linear}The linear budget constraint
  $\probe_\dtime^\p \response \leq 1 $ implies that solution of the ARE (\ref{eq:are}) satisfies ${\kalmancov^*_\dtime}^{-1} (\probe_\dtime,\response) \preceq \bound^{-1}$
  for some symmetric positive definite matrix $\bound^{-1}$.
\end{lemma}

The proof of Lemma \ref{lem:linear} follows straightforwardly using the information Kalman filter formulation \cite{AM79}, and showing that ${\kalmancov^*}^{-1}$ is increasing  in $\response$. Afriat's theorem requires that the constraint
$\probe_\dtime^\p \response \leq 1 $  is active at $\response = \response_\dtime$. This holds in our case
since ${\kalmancov^*}^{-1}$ is increasing  in~$\response$.

To summarize, we can use Afriat's theorem (Theorem \ref{thm:AfriatTheorem}) with $\probe_\dtime$ as the spectrum of $\snoisecov$ and $\response_\dtime$ as the spectrum of $\onoisecov$,  to test a cognitive radar for utility maximization (\ref{eq:radaropt}). Moreover, Afriat's theorem  constructs a set of  utility functions (\ref{eqn:estutility}) that  rationalize the decisions of the radar.

{\bf Summary}: The above approach  of \textit{spectral revealed preferences}, i.e., choosing the probe signal  $\probe$ as the spectrum of $\snoisecov^{-1}$ and the response signal $\response$ as the spectrum of  $\onoisecov$, yields a novel class of methods for detecting utility maximization behavior in a radar. To our knowledge, such methods have not been studied in the context of adaptive sensing or inverse reinforcement learning.
Also, Statement 4 of the theorem is constructive; it yields  explicit utility functions that rationalize the decisions of the cognitive sensor.
From a practical point of view, estimating the enemy sensor's utility function, allows us to predict (in a Bayesian sense) its  future actions  and therefore guard against these actions.

Finally, although we will not elaborate here (due to lack of space), the above framework sets the stage for future work that  addresses the following deeper  questions:
\begin{quote} \textit{How to design cognitive sensors that barely pass Afriat's theorem (or generalization)? Equivalently: How to design  smart (economic rational) sensors that appear dumb?  \\ Are there useful cognitive radars that are not economic rational?}
\end{quote}

\subsection{Detecting Cognitive Radars in a Noisy Setting}  If   the sensor's response $\response_\dtime$ is observed in noise, then violation of Afriat's theorem could be either due to measurement noise or  absence of utility maximization. We will construct  statistical detection tests to decide if the sensor is a utility maximizer and therefore ``cognitive''. Our proposed research builds on stochastic revealed preferences  \cite{FW05,ACD14,AK18},
and our previous works in \cite{KH12,AK17,HNK15}.
 
Our plan proceeds as follows. To highlight the main ideas, 
assume additive measurement errors:
\begin{equation}
  \nresponse_\dtime = \response_\dtime + \anoise_\dtime, \qquad
  \text{ where } \anoise_\dtime \in \mathbb{R}^m\; \text{are  iid  }
	\label{eqn:noisemodel}
\end{equation}
Here $\nresponse_\dtime$ is the noisy measurement of response (action) $\response_\dtime$ of the cognitive sensor.

Given the noisy data set 
$
	\obsdataset= \left\{\left(\probe_\dtime,\obsresponse_\dtime \right): \dtime \in \left\{1,\dots,\horizon \right\}\right\},
$
      how can a statistical test be designed for testing
  utility maximization~\eqref{eq:utilitymaximization} in a dataset due to measurement errors?
Let $H_0$ denote the null hypothesis that the data set $\obsdataset$  satisfies utility maximization. 
Similarly, let $H_1$ denote the alternative hypothesis that the data set does not satisfy utility maximization. 
Then:
\begin{align}
	\text{\bf Type-I errors:}   &\text{\hspace{1.6mm}Reject $H_0$ when $H_0$ is valid.} \nonumber\\
	\text{\bf Type-II errors:}  &\text{\hspace{1.6mm}Accept $H_0$ when $H_0$ is invalid.}
	\label{eqn:hypothesis}
\end{align}

We will design 
statistical tests to detect if an agent is seeking to maximize a utility function. Also statistical tests will be constructed jump changes in the utility function as in our earlier work \cite{AK17}.
These  tests evaluate the probability of the minimum perturbation required for the observed dataset to fall inside the convex polytope specified by the revealed preferences inequalities (\ref{eqn:AfriatFeasibilityTest}).

Finally, we will  derive 
analytical expressions for a lower bound on the false alarm probability.

\subsection{Optimal Probing to Detect Utility Maximization Behavior}

\textit{{What choice of our probe signal yields the smallest Type II error (\ref{eqn:hypothesis}) in detecting if the enemy sensor is a utility maximizer, subject to maintaining the Type I error within a specified bound?}}

\begin{figure}[h] \centering
\resizebox{8cm}{!}{ 
\begin{tikzpicture}[node distance = 4.5cm, auto]
  \node [block3] (BLOCK1) {Estimate Type-II errors and optimize probe};
  \node[block3,right of = BLOCK1](opt){``Our'' State \\ with parameter $\probe_\dtime$ };
    \node [block, below of=BLOCK1,node distance=2.5cm] (BLOCK2) {Radar\\ Controller};
    \node[block, right of=BLOCK2] (grad) {Tracking\\ Algorithm};

    \path [line] (BLOCK1) -- (opt);
    
    \path [line] (opt.east) --++ (1cm,0cm) node[pos=0.25,above]{$\state_k$}  |-   node[pos=0.35,left]{$\obs_k$}    (grad);
    
     \path [line] (BLOCK2.west) --++ (-1cm,0cm)  node[red, pos=0.4,below]{$\response_\dtime$}  |-     node[red, pos=0.7,above]{$\obsresponse_\dtime$}     (BLOCK1.west);
     \path [line] (grad) --   node[pos=0.15,below]{$\belief_k$}    node[red, pos=0.8,above]{$\probe_\dtime$}  (BLOCK2);
     \node[draw, thick, dotted, inner sep=3ex, yshift=-1ex,
     fit=(BLOCK2) (grad)] (box) {};
     \node[fill=white, inner xsep=1ex] at (box.south) {Cognitive Radar};
   \end{tikzpicture}
 }
 \caption{Optimizing the probe waveform to detect cognition in adversary's radar by minimizing the Type II errors subject to constraints in Type 1 errors.
   }
 \label{fig:optimize}
\end{figure}
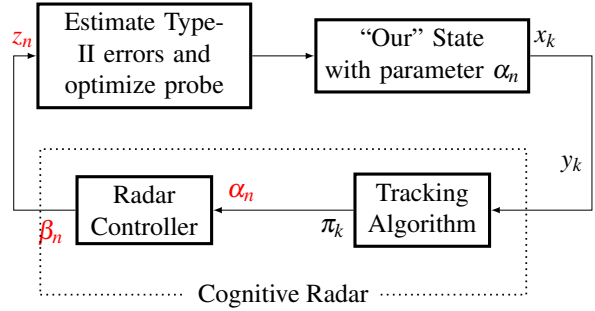

The framework  above guarantees that if we observe the radar response in noise, then the probability of Type-I errors (deciding that the radar is not a utility maximizer  when it is) is less then a threshold $\threshold$ for the decision test. The goal is to enhance  the statistical test  by adaptively optimizing the probe vectors $\Probe=[\probe_1,\probe_2,\dots,\probe_\horizon]$ to reduce the probability of  Type-II errors
(deciding that the radar is cognitive when it is not). 
Our proposed
framework is shown in Figure \ref{fig:optimize}.

We will formulate the optimal probing problem as follows:
Estimate optimal  probe signals $\Probe$ as
{\normalsize 
\begin{align}
  &\argmin_{\Probe\in\reals^{\probedim\times \horizon}_+}J(\Probe) = \underbrace{\prob\big( 
    \text{Error}(\Probe,\Response,\Anoise) 
   > \threshold \big| \{\Probe,\Response(\Probe)\}\in \mathcal{A}\big)}_\text{Probability of Type-II error}. 
\label{eqn: SPSA Objective}
\end{align}
}
Here $\prob(\cdots | \cdot)$ denotes the conditional probability that the statistical test  accepts $H_0$,  given that $H_0$ is false. 
$\Anoise=[\anoise_1,\anoise_2,\dots,\anoise_\horizon]$ is the noise variable defined in
(\ref{eqn:noisemodel}), and $\threshold$ is the significance level.
 The set $\mathcal{A}$ contains all the elements $\{\Probe,\Response(\Probe)\}$, with $\Response(\Probe)=[\response_1,\response_2,\dots,\response_\horizon]$, where  $\{\Probe,\Response\}$ does not satisfy the GARP inequalities of Afriat's Theorem \ref{thm:AfriatTheorem}.

 Since Error$(\cdot)$  is not known explicitly, (\ref{eqn: SPSA Objective}) is a simulation based stochastic optimization problem. To estimate a local minimum value of $J(\Probe)$, we will use stochastic optimization algorithms  such as  the simultaneous perturbation stochastic gradient (SPSA) algorithm \cite{Spa03}.

  \subsection{Bayesian Revealed Preferences and Rational Inattention}
 We now briefly discuss an extension  that invovles Bayesian revealed preferences for detecting cognitive sensors.
Suppose the enemy is a  Bayesian cognitive sensor that
  chooses its  action at each time instant to maximize an \textit{expected} utility function based on the noisy
measurement of an underlying state. Assume that the Bayesian agent is rationally inattentive, that is,
obtaining accurate measurements are expensive. This information acquisition cost affects the observation 
chosen by the agent and hence  affects the action chosen by the agent. We observe the dataset of actions of the Bayesian agent and know
the underlying state (our state).
Our research aims to address the  question:

\begin{quote}
  \textit{\text{How to estimate the utility function and rational inattention cost of the Bayesian enemy?}}
\end{quote}

Bayesian revealed preferences have been studied in the context of human decision making with rational inattention in \cite{CD15,CM15}.
Unlike Afriat's Theorem \ref{thm:AfriatTheorem},  the utility function in the Bayesian set-up is discrete valued; still one can  construct a utility function that  rationalizes the data.
 Rational inattention by Bayesian agents was  pioneered by  Sims \cite{Sim10}. Rational inattention is a form of bounded rationality - the key idea is that a sensor's 
resources  for information acquisition are limited and can be modeled in information
theoretic terms as a Shannon capacity limited communication channel. The action $\action$ and information acquisition policy $\mu$ of a rationally inattentive Bayesian agent are obtained by optimizing the following utility 
\beq \max_\mu \E_y \big\{ \underbrace{\max_a  \E\{ U(x,a) | y,\mu\}}_{\text{Bayesian utility maximization}} - \lambda \big( \underbrace{H(\pi) - H(T(\pi,y,\mu)}_{\begin{array}{c}\text{\footnotesize Rational Inattention cost} \\ \text{\footnotesize (Mutual information)}\end{array}}\big)  \big\} , \quad  \lambda \in \reals_+ \label{eq:ra} \eeq
The first term in (\ref{eq:ra})  is  standard Bayesian utility maximization, namely, choosing action
 $a$ to optimize the conditional expectation of a utility function given noisy measurements $y$. But the measurement $y$ depends on the sensing mode $\mu$ used by the agent. It is here that the second term in (\ref{eq:ra}) comes in.
   The second term is the rational inattention cost: it is the difference in the entropies of the prior and posterior when choosing an attention strategy $\mu$. Clearly the more accurate the posterior,  the lower its entropy and hence the higher the cost. The motivation in behavioral economics by Sims is that human attention is expensive and translating  data into a decision  is constrained by finite Shannon capacity  to process information. In cognitive sensing, the motivation for rational inattention is  that the enemy's sensor incurs a higher measurement cost when it deploys a more accurate sensing mode.

The aim is devise revealed preference methods to detect if  the enemy is a rationally inattentive Bayesian utility maximizer; and then estimate the utility  $U$ in
   (\ref{eq:ra}).
   This involves  the  so called NIAS and NIAC conditions (with suitable modifications using a Bayesian tracker) described in \cite{CD15}, which can be viewed as a Bayesian generalization of GARP in the classical Afriat's theorem.
   One can consider  the more general Renyi entropy and divergence; which allows for more flexibility in modeling sensor information (Renyi entropy is a generalization of Shannon entropy); see \cite{}.
   As another extension, it is worthwhile  consider stopping time problems where the  horizon of decision making is not one step but an integer valued random variable. Optimal stopping time formulations  are widely used in adaptive radars in the context of stochastic control \cite{KD07,KD09,CKH09}. An important example is inverse quickest change detection; by observing the response of a rationally inattentive cognitive sensor and detector, we will devise Bayesian revealed preference methods to estimate the delay penalty and false alarm cost of the enemy's detection algorithm. To  our knowledge, these have not been studied in the literature. Another extension is to estimate the costs involved in multiple stopping time problems \cite{KAB18} given the response of the agent.

   \vspace{0.4cm}
   {\bf Summary and extensions.} The ideas developed in this section  constitute a  principled framework for inverse reinforcement learning of cognitive sensing systems.
 As mentioned earlier, revealed preferences 
addresses a deeper and more fundamental question than classical inverse reinforcement learning: does a utility function exist that rationalizes the
given data (with  signal processing constraints) and if so, estimate it.
The discussion presented above  include several novel features including spectral revealed preferences, detecting utility maximization behavior in noise, optimally probing the enemy, and finally detecting Bayesian rationally inattentive behavior.

To keep the page length short, we have omitted two other ideas that we are currently pursuing: (i) revealed preferences to  detect play from the equilibrium of a potential game \cite{HKA16}; i.e., how to determine if a dataset is consistent with play from the Nash equilibrium of a potential game between the enemy and us. (ii)  revealed preferences when the utility functions are only partially orderable (on a lattice). This arises if the utility function as matrix-valued; e.g.\ covariance matrix of the enemy's Bayesian estimator. This involves estimating preferences on  Hausdorff spaces. 

\section*{Acknowledgments}
The research presented in this chapter was funded by  Air Force Office of Scientific Research grant FA9550-18-1-0007
through the Dynamic Data Driven Application Systems Program.

\bibliographystyle{abbrv}

\bibliography{$HOME/texstuff/styles/bib/vkm}

\end{document}